\documentclass[journal]{IEEEtran}

\usepackage{graphicx}
\usepackage{amsfonts,amsthm,bm}
\usepackage{tabularx}
\usepackage{mathtools}
\usepackage[makeroom]{cancel}
\usepackage{amssymb}
\usepackage{comment}
\usepackage{float}
\usepackage[dvipsnames]{xcolor}
\usepackage{ifpdf}
\usepackage{cite}
\usepackage{hyperref}
\usepackage{enumitem}
\usepackage{algorithm2e}
\usepackage{bbold}
\usepackage{multicol}
\newtheorem{theorem}{Theorem}
\newtheorem{corollary}{Corollary}
\newtheorem{remark}{Remark}
\newtheorem{lemma}{Lemma}
\newtheorem{definition}{Definition}
\newtheorem{proposition}{Proposition}

\begin{document}

\title{A Passivity {Interpretation of Energy-Based} Forced Oscillation Source Location {Methods}}

\author{Samuel Chevalier,~\IEEEmembership{Student Member,~IEEE,}
        Petr Vorobev,~\IEEEmembership{Member,~IEEE,}
        Konstantin Turitsyn,~\IEEEmembership{Member,~IEEE}
\thanks{{This work was supported in part by the Skoltech-MIT Next Generation grant and by the MIT Energy Initiative Seed Fund Program.}

{S. Chevalier is with Department of Mechanical Engineering, Massachusetts Institute of Technology. E-mail: schev@mit.edu}

{P. Vorobev is with the Skolkovo Institute of Science and Technology (Skoltech), Moscow, Russia. E-mail: p.vorobev@skoltech.ru}

{K. Turitsyn was with Department of Mechanical Engineering, Massachusetts Institute of Technology. E-mail: turitsyn@gmail.com}}}

\maketitle

\begin{abstract}
This paper develops a systematic framework for analyzing how low frequency forced oscillations propagate in electric power systems. Using this framework, the paper shows how to mathematically justify the so-called Dissipating Energy Flow (DEF) forced oscillation source location technique. The DEF's specific deficiencies are pinpointed{, and its underlying energy function is analyzed via incremental passivity theory. This analysis is then used to prove that there exists no passivity transformation (i.e. quadratic energy function) which can simultaneously render all components of a lossy classical power system passive. The paper goes on to develop a simulation-free algorithm for predicting the performance of the DEF method in a generalized power system, and it analyzes the passivity of three non-classical load and generation components. The proposed propagation framework and performance algorithm are both tested and illustrated} on the IEEE 39-bus New England system {and the WECC 179-bus system}.
\end{abstract}


\IEEEpeerreviewmaketitle


\section{Introduction}\label{Introduction}

\IEEEPARstart{F}{orced} oscillations (FOs) are still a problematic reality in modern electric power systems. Caused by extraneous periodic perturbations, FOs can compromise system security, degrade system performance, and resonantly excite poorly damped interarea modes~\cite{Vanfretti:2012,NERC:2017,Nezam:2016}. Despite widescale deployment of Phasor Measurement Units (PMUs) across the US high voltage transmission network, locating the sources of these FOs remains a challenging task due to their sporadic nature, speed of propagation, and inability to be predicted by the system operators' dynamical models. Since the most effective way for dealing with a FO is to locate the source and disconnect it from service~\cite{Maslennikov:2017}, an effective source location technique is an indispensable smart grid application.

Of the many source location techniques currently available in the academic marketplace~\cite{Wang:2017}, the so-called Dissipating Energy Flow (DEF) method has enjoyed some of the most successful testing results, both in simulation environments~\cite{Maslennikov:2017} and in real-time applications~\cite{Maslennikov:2019}. The method was originally developed by Chen et al.~\cite{Chen:2013}, but its underlying mathematics leverage the Lyapunov functions from~\cite{Tsolas:1985}. Despite its success, when its underlying modeling assumptions are violated, the method may perform poorly~\cite{Chen:2017,Chevalier:2018,Chevalier:2019P}. Additionally, the method has lacked a generalized framework which is capable of providing a system-wide justification for its usage.

{The primary challenge to reliable DEF performance is the contribution of dissipating energy from non-FO sources, such as lossy transmission lines, lossy or negatively damped loads, and generator dynamics dominated by non-passive controllers. An actual example of this contribution may be found in Fig. 1 of~\cite{Chevalier:2019P}. If these contributions are large enough, the FO source can appear to be a dissipating energy sink and the DEF method can fail.} Despite its inadequacies, the DEF's excellent performance in real-time application at ISO New England strongly implies that further research should be performed in order to more systematically characterize the method. Shortcomings of the DEF method have been analyzed in~\cite{Chen:2017}, and~\cite{Chevalier:2019P} has recommended using passivity theory to interpret the method from a new mathematical perspective, but no {theoretical methods have been devised for testing how the DEF method will perform in an arbitrary network. Such testing is essential in order to ensure that the DEF can perform adequately in new environments, such as in microgrids where ``R/X" line ratios are high and voltage control is fast, or in networks that have particularly resistive load pockets. To make such predictions,} a systematic framework is needed in order to thoroughly study the DEF.

{Accordingly, this paper develops a framework which is suitable for analyzing the DEF method via passivity theory in the context of a full-scale system. Initially, this framework is used to analyze a lossy classical power system, but the methods are then generalized to include arbitrary system models; these generalized methods are thus capable of investigating all of the problematic effects presented by lines, loads and generator controllers. The primary contributions of this paper are as follows:}

\begin{enumerate}

\item Using the Frequency Response Function (FRF) analysis proposed in~\cite{Chevalier:2018}, we leverage a variety of tools from AC circuit theory in order to develop a linearized framework for analyzing oscillation propagation at the system level.

\item We subsequently use the proposed framework, along with the passivity observations from~\cite{Chevalier:2019P}, to theoretically justify the DEF method and show that there exists no other quadratic passivity transformation which will render all components of the entire network passive in a classical power system.

\item {A simulation-free algorithm is developed which predicts the performance of the DEF method in a generalized power system.}

\item {The proposed passivity transformation is used to analytically investigate the dissipating energy flow properties of three common, yet non-classical, grid components.}

\end{enumerate}

{The remainder of this paper is structured as follows. In Section \ref{Network Model}, we derive the linearized propagation framework and show how quadratic energy is conserved in the system. In Section \ref{Energy Function Analysis}, we extend the energy function analysis and prove that there is no quadratic energy function which will render all components of a classical power system passive. We then leverage the proposed framework in Section \ref{Practical Applications} in order to develop an algorithm which analytically predicts the performance of the DEF method for a generalized power system. We present illustrative and supportive test results in Section \ref{Test Results} and offer concluding remarks in Section \ref{Conclusion}.}


\section{Perturbative Network Model Derivation}\label{Network Model}
In this section, we introduce a linearized network model which is particularly useful for analyzing FO propagation in power systems. We refer to it as a ``perturbative" model since it captures the network's response to small\footnote{{The framework presented in this paper is valid when FOs are sufficiently ``small", such that quadratic nonlinearities may be neglected.}} perturbations.
\subsection{Complex Admittance Matrices}
\begin{figure}[H]
\centering
\includegraphics[scale=1.3]{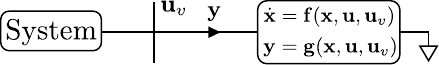}
\caption{\label{fig: DAE_Model} DAE modeled component tied to a larger power system.}
\end{figure}

We consider a power system component, shown in Fig. \ref{fig: DAE_Model}, whose dynamics are governed according to the DAE set
\begin{subequations}\label{eq: nonlin_sys}
\begin{align}
\dot{{\bf x}} & ={\bf f}({\bf x},{\bf u},{\bf u}_v)\\
{\bf y} & ={\bf g}({\bf x},{\bf u},{\bf u}_v),
\end{align}
\end{subequations}
where inputs ${\bf u}_{v}=\left[V_{r},\,V_{i}\right]$ and outputs ${\bf y}=\left[I_{r},\,I_{i}\right]$ are vectors of real and imaginary voltages and currents, respectively.
We linearize this model around a steady state operating point to linearly relate the voltage and current perturbations:
\begin{subequations}\label{eq: state_space}
\begin{align}
\Delta\dot{{\bf x}} & =A\Delta{\bf x}+B\Delta{\bf u}_{v}\\
\Delta{\bf y} & =C\Delta{\bf x}+D\Delta{\bf u}_{v}.
\end{align}
\end{subequations}
Assuming (\ref{eq: state_space}) is BIBO stable, its Fourier transform admits the Frequency Response Function (FRF) of the component~\cite{Chevalier:2018}:
\begin{align}
\tilde{{\bf y}}=\underbrace{\left(C(j\Omega{\mathbb 1}-A)^{-1}B+D\right)}_{\mathcal{Y}(j\Omega)}\tilde{{\bf u}}_{v},\label{eq: Y(jOmega)}
\end{align}
where $\Omega$ is the angular frequency of the input and output signals, and $\mathcal{Y}\equiv\mathcal{Y}(j\Omega)\!\in\!{\mathbb C}^{2\times2}$ is referred to as the admittance matrix relating voltage $\tilde{{\bf u}}_{v}\!\!\in\!\!{\mathbb C}^{2\times1}$ and current $\tilde{{\bf y}}\!\!\in\!\!{\mathbb C}^{2\times1}$ perturbations. These perturbations can be given in rectangular (${\tilde V}_r$, ${\tilde V}_i$) or polar (${\tilde {\rm V}}$, ${\tilde \theta}$) coordinates depending on convenience. In this paper, we will primarily consider the effects of the FRF matrix at the relevant forcing frequency $\Omega_d$ of the FO. {When referring to $\mathcal Y$, we will use the terms FRF and admittance interchangeably, depending on the context.}

\subsection{Network Modeling}
Consider a power system network whose graph ${\mathcal G}({\mathcal V},{\mathcal E})$ has edge set $\mathcal{E}$, $|\mathcal{E}|=m$, vertex set $\mathcal{V}$, $|\mathcal{V}|=n$, and directed nodal incidence matrix $E\in{\mathbb R}^{m\times n}$~\cite{Amini:2016}. {In power system modeling, it is standard practice to use the incidence matrix to build the nodal admittance matrix $Y_{\rm bus}$ from the primitive admittance matrix $Y_p$ via $Y_{\rm bus}=E^{\dagger}Y_pE$~\cite{Nagappan:1970}. This procedure is possible because the admittances in $Y_p$ are complex values $\in {\mathbb C}^1$. In the following analysis, all admittances in the form of (\ref{eq: Y(jOmega)}) must be expressed in $2\times 2$ matrix form (i.e. $\mathcal Y\in {\mathbb C}^{2\times 2}$). Accordingly, we need an incidence matrix which can relate $2\times 2$, rather than $1\times 1$, admittance values. We thus build a so-called ``augmented"} incidence matrix $E_a\in{\mathbb R}^{2m\times 2n}$. This matrix is constructed by taking $E$ and replacing all values of 1 with the $2\times2$ identity matrix $\mathbb{1}_2$ and all values of 0 with a $2\times2$ zero matrix ${\bf 0}$:
\begin{align}\label{eq: Ea}
E=\left[\!\begin{array}{ccc}
1 & -1 & \!\!\!\cdots\\
0 & 1\vspace{-0.15cm} \\
 \vdots & & \!\!\!\ddots
\end{array}\!\right]\,\Leftrightarrow\,\,\,E_{a}=\left[\!\begin{array}{ccc}
\mathbb{1}_2 & -\mathbb{1}_2 & \!\!\!\cdots\\
{\bf 0} & \mathbb{1}_2\vspace{-0.15cm}\\
 \vdots & & \!\!\!\ddots
\end{array}\!\right].
\end{align}
Considering voltage and current perturbation phasors such as $\tilde{{\bf u}}_{v}$ and $\tilde{{\bf y}}$ from (\ref{eq: Y(jOmega)}), we define the vector ${\bf V}_{b}\in{\mathbb C}^{2n\times 1}$ as the vector of rectangular bus voltage perturbation phasors, and  we define the vector ${\bf I}_{l}\in{\mathbb C}^{2m\times 1}$ as the vector of rectangular line current perturbation phasors, where the convention of the positive line current flows agrees with the direction of the augmented incidence matrix:
\begin{tabularx}{\linewidth}{@{}XX@{}}
\begin{equation}
   {\bf V}_{b} =\left[\!\begin{array}{c}
    \tilde{V}_{r,1}\\
    \tilde{V}_{i,1}\\
    \vdots\\
    \tilde{V}_{r,n}\\
    \tilde{V}_{i,n}
    \end{array}\!\right],
\end{equation}
&
\begin{equation}
  {\bf I}_{l} =\left[\!\begin{array}{c}
    \tilde{I}_{r,1}\\
    \tilde{I}_{i,1}\\
    \vdots\\
    \tilde{I}_{r,m}\\
    \tilde{I}_{i,m}
    \end{array}\!\right].
\end{equation}
\end{tabularx}
Admittance matrices $\mathcal{Y}_{l,1},...,\mathcal{Y}_{l,m}$, $\mathcal{Y}_{l,i}\in\mathbb{R}^{2\times2}$, associated with the $m$ transmission lines, are placed diagonally in the matrix ${\mathcal Y}_L \in {\mathbb R}^{2m\times2m}$ such that
\begin{align}\label{eq: YL_mat}
\mathcal{Y}_{L} & =\left[\begin{array}{ccc}
\mathcal{Y}_{l,1} &  & {\bf 0}\vspace{-0.15cm}\\
 & \!\!\!\ddots\vspace{-0.15cm}\!\!\!\\
{\bf 0} &  & \mathcal{Y}_{l,m}
\end{array}\right].
\end{align}
{Transformers with off-nominal tap ratios, such as tap changing transformers, are discussed in Appendix \ref{AppA}.} Line current and bus voltage perturbations obey Ohm's law:
\begin{align}
\mathcal{Y}_{L}E_a{\bf V}_{b} & ={\bf I}_{l}.
\end{align}
Admittance matrices $\mathcal{Y}_{s,1},...,\mathcal{Y}_{s,n}$, $\mathcal{Y}_{s,i}\in\mathbb{C}^{2\times2}$, associated with shunt elements at each of the $n$ buses, are placed diagonally in the matrix ${\mathcal Y}_S \in {\mathbb C}^{2n\times2n}$, such that
\begin{align}\label{eq: YS_mat}
\mathcal{Y}_{S}=\left[\begin{array}{ccc}
\mathcal{Y}_{s,1}\!\!\! &  & \!\!\!{\bf 0}\vspace{-0.15cm}\\
 & \ddots\vspace{-0.15cm}\\
{\bf 0}\!\!\! &  & \!\!\!\mathcal{Y}_{s,n}
\end{array}\right].
\end{align}
These shunt admittances are not simply capacitors or inductors; they can represent generators, loads, or any other terminal element in the system and can be constructed via (\ref{eq: Y(jOmega)}). If multiple elements are connected in parallel, such as a generator and its station load, their admittances can be modeled independently and summed to compute the aggregate shunt admittance. The shunt matrix ${\mathcal Y}_S$ may be used to compute the shunt current injections\footnote{Shunt currents flowing out of the circuit to ground are defined as positive.} ${\bf I}_s\!\in\!{\mathbb C}^{2n\times1}$ via
\begin{align}
{\bf I}_{s}=\mathcal{Y}_{S}{\bf V}_{b}.
\end{align}
As outlined in~\cite{Chevalier:2018}, when analyzing a network with this representation, FOs show up like current sources at their respective source buses. For a system experiencing a \textit{single} FO, there will be a \textit{single} current source {${\bm{\mathcal I}}\in{\mathbb C}^{2\times 1}$} driving the entire network. We define a sparse FO vector of current injections ${\bf J}\in\mathbb{C}^{2n\times1}$ {whose structure will take the form}
\begin{align}
{{\bf J}=[{\bf 0},\,\ldots{\bf 0},\,{\bm{\mathcal I}},\,{\bf 0},\,\ldots{\bf 0}]^{\top}.}
\end{align}
{If bus $k$ is the source of the FO, then ${\bm{\mathcal I}}$ will be located in elements ${\bf J}_{2k-1}$ and ${\bf J}_{2k}$.} Source injections in ${\bf J}$ obey the same current convention as ${\bf I}_s$. The network obeys KCL, i.e., all nodal currents sum to 0:
\begin{align}
{\bf J}+{\bf I}_{s}+E_a^{\dagger}{\bf I}_{l}={\bf 0}.
\end{align}
We define ${\bf I}_I\equiv E_a^{\dagger}{\bf I}_{l}$ to be the aggregate current injection at each node: it represents the sum of the source current injection at each bus plus the shunt current flowing to ground. Via conservation of current at each bus, we have
\begin{subequations}
\begin{align}
-{\bf J} & ={\bf I}_{I}+{\bf I}_{s}\\
 & =E_a^{\dagger}{\bf I}_{l}+\mathcal{Y}_{S}{\bf V}_{b}\\
 & =\left(E_a^{\dagger}\mathcal{Y}_{L}E_a+\mathcal{Y}_{S}\right){\bf V}_{b}.\label{eq: J=Yfull_V}
\end{align}
\end{subequations}
The block-Hermitian dynamic nodal admittance (or augmented dynamic Y-bus) matrix $\mathcal{Y}_{B}\in{\mathbb C}^{2n\times2n}$ is this defined to be
\begin{align}\label{eq: Yb_Model}
\mathcal{Y}_{B}=E_a^{\dagger}\mathcal{Y}_{L}E_a+\mathcal{Y}_{S}.
\end{align}
Assuming there is a single FO in the system, it is instructive to rewrite (\ref{eq: J=Yfull_V}) with partitioned matrices and vectors, where the system has been renumbered such that the source bus is bus 1, and the current injection has a value of ${\bm{\mathcal I}}\in{\mathbb C}^{2\times1}$:
\begin{subequations}\label{eq: J=YV_full}
\begin{align}\label{eq: J=YV}
-{\bf J} & =\mathcal{Y}_{B}{\bf V}_{b}\\
\left[\begin{array}{c}
-{\bm{\mathcal I}}\\
\hline {\bf 0}
\end{array}\right] & =\left[\begin{array}{c|c}
\mathcal{Y}_{B1} & \mathcal{Y}_{B2}\\
\hline \mathcal{Y}_{B3} & \mathcal{Y}_{B4}
\end{array}\right]\left[\begin{array}{c}
{\bf V}_{s}\\
\hline {\bf V}_{ns}
\end{array}\right],\label{eq: Net_Demcop}
\end{align}
\end{subequations}
where ${\bf V}_{s}\!\in\!{\mathbb C}^{2\times1}$ represents voltage perturbations at the source bus, ${\bf V}_{ns}\!\in\!{\mathbb C}^{(2n-2)\times1}$ represents voltage perturbations at all other buses, and ${\bf V}_b={\bf V}_{s}^{\frown}{\bf V}_{ns}$. While ${\bm{\mathcal I}}$ represents the true source current injection, we may also define ${\bm{\mathcal I}}'$ as the sum of the source current at the source bus plus its shunt injection:
\begin{align}
{\bm{\mathcal I}}'={\bm{\mathcal I}} + {\mathcal Y}_{s,1}{\bf V}_s.
\end{align}
Correspondingly, we say that ${\mathcal Y}_{B1}'$ contains no shunt element, and ${\bm{\mathcal I}}'$ is the current \textit{directly measured at the source bus flowing into the network}; we note that it is equal to the first two elements of ${\bf I}_I$. We now restate (\ref{eq: J=YV})-(\ref{eq: Net_Demcop}) with this update:
\begin{subequations}\label{eq: Net_Demcop_both}
\begin{align}\label{eq: J=YV'}
-{\bf J}' & =\mathcal{Y}_{B}'{\bf V}_{b}\\
\left[\begin{array}{c}
-{\bm{\mathcal I}}'\\
\hline {\bf 0}
\end{array}\right] & =\left[\begin{array}{c|c}
\mathcal{Y}_{B1}' & \mathcal{Y}_{B2}\\
\hline \mathcal{Y}_{B3} & \mathcal{Y}_{B4}
\end{array}\right]\left[\begin{array}{c}
{\bf V}_{s}\\
\hline {\bf V}_{ns}
\end{array}\right].\label{eq: Net_Demcop'}
\end{align}
\end{subequations}
A simple Kron reduction can be performed in order to determine the effective admittance ``seen" by the current source:
\begin{align}
-{\bm{\mathcal I}}' & =\underbrace{\left(\mathcal{Y}_{B1}'-\mathcal{Y}_{B2}\mathcal{Y}_{B4}^{-1}\mathcal{Y}_{B3}\right)}_{\mathcal{Y}_{N}}{\bf V}_{s},\label{eq: YN1}
\end{align}
where $\mathcal{Y}_{N}\!\in\!{\mathbb C}^{2\times2}$ is a $2\times2$ complex aggregate network admittance matrix and ${\bf V}_{s}$ is the resulting voltage caused by the current injection ${\bm{\mathcal I}}'$ interacting with the aggregate network dynamics codified in $\mathcal{Y}_{N}$. In this model, since voltage perturbations are considered a response to rouge current injections, it is helpful to rewrite (\ref{eq: YN1}) with voltage as a function of current:
\begin{align}
{\bf V}_{s}=-\mathcal{Z}_{N}{\bm{\mathcal I}}',\label{eq: ZN}
\end{align}
where $\mathcal{Z}_{N}\!=\!\mathcal{Y}_{N}^{-1}$ is the aggregate network impedance. In other words, the current injection ${\bm{\mathcal I}}'$ \textit{gives rise} to the network voltages, and the vector ${\bf V}_b$ in (\ref{eq: J=YV'}) is not arbitrary: the Kron reduction of (\ref{eq: YN1}) is only meaningful when ${\bf V}_{b}$ acts as a solution to the linear system $-{\bf J}'=\mathcal{Y}_{B}'{\bf V}_{b}$, i.e. ${\bf V}_{b}=-(\mathcal{Y}_{B}')^{-1}{\bf J}'$.
\begin{definition}\label{def: DWE}
We refer to the admittance matrix ${\mathcal Y}_N$ of (\ref{eq: YN1}) as the system's dynamic Ward equivalent (\textbf{DWE}) admittance.
\end{definition}
\subsection{Quadratic Energy Considerations}
As with any network which obeys Kirchhoff's laws, Tellegen's theorem is also obeyed: the sum of the products of branch (including shunt branches) potential differences and branch flows is equal to 0. Accordingly,
\begin{subequations}
\begin{align}
0 & =\left(E_{a}{\bf V}_{b}\right)^{\dagger}{\bf I}_{l}+{\bf V}_{b}^{\dagger}\left({\bf I}_{s}+{\bf J}\right)\label{eq: Tel1}\\
 & ={\bf V}_{b}^{\dagger}\left(E_{a}^{\dagger}{\bf I}_{l}+{\bf I}_{s}+{\bf J}\right)\label{eq: Tel2}\\
 & ={\bf V}_{b}^{\dagger}\left(\mathcal{Y}_{B}-\mathcal{Y}_{B}\right){\bf V}_{b},\label{eq: Tel3}
\end{align}
\end{subequations}
where (\ref{eq: Tel1}) is the statement of Tellegen's theorem, (\ref{eq: Tel2}) is the conservation of current, and (\ref{eq: Tel3}) is the resulting proof. As a consequence of this theorem, there exist a family of quadratic functionals for which conservation laws can be formulated. An obvious one is the ``real power", $\rm{Re}\{{\bf V}{\bf I}^{\dagger}\}$, that is consumed only on the elements with positive resistance. In the context of FOs, an alternative interpretation of the conservation of power can be acquired by manipulating (\ref{eq: Net_Demcop'}) in order to define another (arbitrary) type of quadratic power\footnote{The term ``quadratic power" is used since we are multiplying voltages and currents, but the quadratic quantity doesn't necessarily have the interpretation of physical power. It can also be interpreted as an ``energy function". Quadratic energy and quadratic power are therefore used interchangeably.}. The key observation is that this new quadratic power will be conserved throughout the network. To show why, we consider matrix ${\mathcal Q}\in{\mathbb C}^{2n\times2n}$ with block diagonal sub-matrices ${\mathcal Q}_b\in{\mathbb C}^{2\times2}$:
\begin{align}\label{eq: Q}
{\mathcal Q} & =\left[\begin{array}{ccc}
{\mathcal Q}_b\!\!\! &  & \!\!\!{\bf 0}\vspace{-0.15cm}\\
 & \ddots\vspace{-0.15cm}\\
{\bf 0}\!\!\! &  & \!\!\!{\mathcal Q}_b
\end{array}\right].
\end{align}
We now left multiply (\ref{eq: J=YV'}) by ${\bf V}_{b}^{\dagger}{{\mathcal Q}}$, which represents the application of a quadratic energy function:
\begin{subequations}
\begin{align}
-{\bf V}_{b}^{\dagger}{{\mathcal Q}}{\bf J}' & ={\bf V}_{b}^{\dagger}{{\mathcal Q}}\mathcal{Y}_{B}'{\bf V}_{b}\\
-{\bf V}_{s}^{\dagger}\mathcal{Q}_{b}{\bm{\mathcal I}}' & ={\bf V}_{b}^{\dagger}\left(\mathcal{Q}E_a^{\dagger}\mathcal{Y}_{L}E_a\right){\bf V}_{b}+{\bf V}_{b}^{\dagger}\left(\mathcal{Q}\mathcal{Y}_{S}'\right){\bf V}_{b}.\label{eq: Network_Passivity}
\end{align}
\end{subequations}
Different choices for matrix ${{\mathcal Q}}_{b}$ correspond to different energy function applications, but in each case, the quadratic quantity is conserved. For example, if ${\mathcal Q}_b$ is chosen such that 
\begin{align}\label{eq: DEF_Qb}
{\mathcal Q}_b=\left[\!\begin{array}{cc}
0 \!& -\frac{1}{j\Omega}\\
\frac{1}{j\Omega} \!& 0
\end{array}\!\right],
\end{align}
then the associated energy function corresponds to the DEF method~\cite{Chevalier:2019P}. Under DEF assumptions, lines and loads are rendered lossless, i.e $(\mathcal{Q}_b\mathcal{Y}) + (\mathcal{Q}_b\mathcal{Y})^\dagger={\bf 0}$. Thus, in taking the real part, (\ref{eq: Network_Passivity}) simplifies to
\begin{align}\label{eq: DEF_f}
\underbrace{{\rm Re}\{{\bf V}_{s}^{\dagger}\mathcal{Q}_{b}{\bm{\mathcal I}}'\}}_{\text{Source Energy}}+\sum_{i=2}^{n}\underbrace{{\rm Re}\{{\bf V}_{ns,i}^{\dagger}\left(\mathcal{Q}_{b}\mathcal{Y}_{s,i}\right){\bf V}_{ns,i}\}}_{\text{Generator Damping Contributions}}=0,
\end{align}
where ${\bf V}_{s}$ and ${\bf V}_{ns,i}$ are the source and $i^{\rm th}$ non-source bus voltage perturbation vectors, respectively. The formulation of (\ref{eq: DEF_f}) further clarifies the DEF's functionality: since all the damping energy consumed by generators is positive~\cite{Chevalier:2019P}, the source energy is necessarily negative and can be traced back to the single, negative source. The DEF technique, therefore, is based on tracking a particular type of quadratic power in the network. When constructing the system's quadratic energy function, we are not restricted to choosing just a ${\mathcal Q}$ matrix. We may also introduce matrix ${\mathcal P}$ whose structure matches ${\mathcal Q}$. For example, we may set ${\bf U}_{b}={\mathcal P}^{-1}{\bf V}_{b}$, left multiply (\ref{eq: J=YV'}) by ${\bf U}_{b}^{\dagger}{{\mathcal Q}}$, and insert a ${\mathcal P}{\mathcal P}^{-1}$ term. Updating (\ref{eq: DEF_f}) yields:
\begin{align}\label{eq: DEF_f2}
\underbrace{{\rm Re}\{{\bf U}_{s}^{\dagger}\mathcal{Q}_{b}{\bm{\mathcal I}}'\}}_{\text{Source Energy}}+\sum_{i=2}^{n}\underbrace{\textbf{{\rm Re}\{}{\bf U}_{ns,i}^{\dagger}\left(\mathcal{Q}_{b}\mathcal{Y}_{s,i}\mathcal{P}_{b}\right){\bf U}_{ns,i}\}}_{\text{Generator Damping Contributions}}=0.
\end{align}


\section{Energy Function Analysis}\label{Energy Function Analysis}
The DEF method can be interpreted as the application of a specific quadratic energy function to all elements of a network. Reference~\cite{Chevalier:2019P}, which reviews relevant passivity concepts, explains how this quadratic energy function can be interpreted as a matrix transformation (called a ``passivity transformation") which attempts to render system elements incrementally passive. Generally, a passive component is one which can only dissipate and store, but not produce, physical power. To avoid confusion, we point out that this paper discusses the so-called \textit{incremental passivity}, which refers to the passive nature of a system's incremental change from its equilibrium, {as defined by~\cite[eq. (4.159)]{Schaft:2000}}. This is also known as ``shifted passivity". We stress that physical passivity of an element does not imply incremental passivity {when some transformation, such as (\ref{eq: Q}), has been applied}. In the remainder of the paper, the term passivity always refers to incremental passivity.

Reference~\cite{Chevalier:2019P} {further} shows that the DEF energy function is inadequate for lossy network elements. While it may be tempting to develop a new passivity transformation which is suitable for lossy networks, in this section, we use passivity theory to prove that no constant quadratic energy function exists for the classical model of a multimachine power system. {For coherency, we first offer the following clarification of how we use the terms ``lossy" and ``lossless" in this paper.}

\begin{definition}
{In keeping with the conventional nomenclature for both power systems and passivity literature, we offer the following definitions for lossy and lossless:
\begin{itemize}
    \item {\bf Lossy}: A lossy element is an element which contains resistance; a lossy system is a system which contains resistive transmission lines.
     \item {\bf Lossless}\footnote{{While ``lossless" certainly can also refer to a transmission network with purely reactive lines, we do not invoke that definition anywhere in this paper.}}: The term lossless describes a FRF whose Hermitian part is $\bf 0$. If ${\mathcal Y} + {\mathcal Y}^{\dagger}={\bf 0}$, ${\mathcal Y}$ is said to be lossless; if ${\bf M}{\mathcal Y} + ({\bf M}{\mathcal Y})^{\dagger}={\bf 0}$, ${\bf M}{\mathcal Y}$ is said to be lossless.
\end{itemize}}
\end{definition}

\subsection{Basis Matrices}
To aide in the energy function analysis and inference techniques, we define a useful set of basis matrices.
\begin{definition}
We define orthogonal ($A^{-1}\!=\!A^{\dagger}$) basis matrices
\begin{align*}
\overline{T}_{1} & =\left[\begin{array}{cc}
1 & 0\\
0 & 1
\end{array}\right], &
\overline{T}_{2} & =\left[\begin{array}{cc}
1 & 0\\
0 & -1
\end{array}\right],\\
\overline{T}_{3} & =\left[\begin{array}{cc}
0 & 1\\
1 & 0
\end{array}\right], &
\overline{T}_{4} & =\left[\begin{array}{cc}
0 & -1\\
1 & 0
\end{array}\right],
\end{align*}
and the set $\mathbb{T}=\{\overline{T}_{1},\overline{T}_{2},\overline{T}_{3},\overline{T}_{4}\}$. Set $\mathbb{T}$ spans region $\mathbb{R}^{2\times2}$:
\begin{align}
{{\rm span}\left(\mathbb{T}\right)=\left\{ \left.\sum_{i=1}^{4}\lambda_{i}\overline{T}_{i}\right|\overline{T}_{i}\in\mathbb{T},\,\lambda_{i}\in\mathbb{R}^{1}\right\}=\mathbb{R}^{2\times2}.}
\end{align}
\end{definition}
\begin{lemma}\label{lemma1}
There exists no non-singular matrix $\Gamma\in\mathbb{C}^{2\times2}$ for which, simultaneously,
\begin{align}
{\rm Re}\left\{ {{\bf v}}^{\dagger}\left(\overline{T}_{i}\Gamma\right){{\bf v}}\right\}  & \ge0,\;\forall{{\bf v}}\in\mathbb{C}^{2\times1},\,\forall i\in\{1\}\label{eq: G_ineq}\\
{\rm Re}\left\{{{\bf v}}^{\dagger}\left(\overline{T}_{i}\Gamma\right){{\bf v}}\right\}  & =0,\;\forall{{\bf v}}\in\mathbb{C}^{2\times1},\,\forall i\in\{2,3,4\}.\label{eq: G_eq}
\end{align}
\begin{proof}
We write $\Gamma$ as the sum of its diagonal ($\Gamma_{d}$) and off-diagonal ($\Gamma_{o}$) component matrices: $\Gamma=\Gamma_{d}+\Gamma_{o}$. Since ${\rm Re}\{ {{\bf v}}^{\dagger}(\overline{T}_{i}\Gamma){{\bf v}}\}=0,\;\forall{{\bf v}}\in\mathbb{C}^{2\times2}$ is equivalent to stating that $\overline{T}_{i}\Gamma +(\overline{T}_{i}\Gamma)^{\dagger}={\bf 0}$, the constraints on $\Gamma$ caused by ${\overline T}_{2}$, ${\overline T}_{3}$, and ${\overline T}_{4}$ from (\ref{eq: G_eq}) may be stated as
\begin{align}
&& {\overline T}_{2}&\;\rightarrow &       \Gamma_d&=-\Gamma_d^{\dagger},    &     \Gamma_o&=\Gamma_o^{\dagger},   &&\\
&& {\overline T}_{3}&\;\rightarrow &      K\Gamma_d&=-\Gamma_d^{\dagger}K,   &    K\Gamma_o&=-\Gamma_o^{\dagger}K, &&\\
&& {\overline T}_{4}&\;\rightarrow &      K\Gamma_d&= \Gamma_d^{\dagger}K,   &    K\Gamma_o&=-\Gamma_o^{\dagger}K, &&
\end{align}
where $K\equiv{\overline T}_{3}$ is defined to be the reversal matrix in~\cite{Horn:1990}. Accordingly, $\Gamma_d$ must be simultaneously skew-Hermitian, skew-perhermitian and perhermitian, respectively~\cite{Horn:1990}. Necessarily, $\Gamma_{d}={\bf 0}$. The matrix $\Gamma_o$ must be simultaneously Hermitian and skew-perhermitian. Necessarily, $j\beta{\overline T}_{4},\,\beta\!\in\mathbb{R}^{1}$, is the only matrix which fits this description. We define
\begin{align}\label{eq: Gam_star}
\Gamma^{\star}&=j\beta{\overline T}_{4}
\end{align}
as the only matrix which uniformly satisfies (\ref{eq: G_eq}). We apply $\Gamma=\Gamma^{\star}$ to (\ref{eq: G_ineq}) and consider the eigenvalues of the matrix ${\overline T}_{1}\Gamma^{\star}+\left({\overline T}_{1}\Gamma^{\star}\right)^{\dagger}=2\beta j{\overline T}_{4}$:
\begin{align}
{\rm det}\left[2\beta j{\overline T}_{4}-{\rm diag}\{{\lambda}\} \right] & =\lambda^{2}-4\beta^{2},\\
\lambda & =\pm2\beta\label{ref: eig_vals_R}
\end{align}
which violates (\ref{eq: G_ineq}). Since ${\overline T}_{1}\Gamma^{\star}\!+\!({\overline T}_{1}\Gamma^{\star})^{\dagger}$ is an indefinite matrix but $\Gamma^{\star}$ is the only matrix which satisfies (\ref{eq: G_eq}), the theorem has been proved.
\end{proof}
\end{lemma}
\begin{corollary}
$\Gamma^{\star}$ is a solution to $P\Gamma^{\star}+(P\Gamma^{\star})^{\dagger}={\bf 0}$ for any matrix $P$ which may be written as $P=\sum_{i=2}^4\alpha_i{\overline T}_i,\,\alpha_i\!\in\!{\mathbb R}^1$.
\end{corollary}
\begin{corollary}
The results of Lemma \ref{lemma1} stand if ${\overline T}_{i}$ is right multiplied by transformation matrix ${\bf M}$ instead of left multiplied by transformation matrix $\Gamma$. There exists no non-singular matrix ${\bf M}\in\mathbb{C}^{2\times2}$ for which the following simultaneously hold:
\begin{align}
{\rm Re}\left\{ {{\bf v}}^{\dagger}\left({\bf M}\overline{T}_{i}\right){{\bf v}}\right\}  & \ge0,\;\forall{{\bf v}}\in\mathbb{C}^{2\times2},\,\forall i\in\{1\}\label{eq: M_ineq2}\\
{\rm Re}\left\{ {{\bf v}}^{\dagger}\left({\bf M}\overline{T}_{i}\right){{\bf v}}\right\}  & =0,\;\forall{{\bf v}}\in\mathbb{C}^{2\times2},\,\forall i\in\{2,3,4\}\label{eq: M_eq2}
\end{align}
\end{corollary}
\begin{corollary}\label{col_M_G}
By employing both non-singular matrices $\Gamma\in\mathbb{C}^{2\times2}$ and ${\bf M}\in\mathbb{C}^{2\times2}$, the solution to
\begin{align}
{\rm Re}\left\{ {{\bf v}}^{\dagger}\left({\bf M}\overline{T}_{i}\Gamma\right){{\bf v}}\right\}  & =0,\;\forall{{\bf v}}\in\mathbb{C}^{2\times1},\,\forall i\in\{2,3,4\}\label{eq: M_eq3}
\end{align}
must take the form $\Gamma=(j\beta\overline{T}_{4}){\bf M}^{\dagger}$ for any ${\bf M}\in{\mathbb C}^{2\times2}$. This may be seen via the following manipulation:
\begin{subequations}
\begin{align}
{\bf 0} & ={\bf M}\overline{T}_{i}\Gamma+({\bf M}\overline{T}_{i}\Gamma)^{\dagger}, & \forall i & \in\{2,3,4\}\\
 & =\overline{T}_{i}(\Gamma {\bf M}^{\dagger^{-1}})+({\bf M}^{-1}\Gamma^{\dagger})\overline{T}_{i}^{\dagger}, & \forall i & \in\{2,3,4\}{.}\label{eq: NT+NT'}
\end{align}
\end{subequations}
Since (\ref{eq: NT+NT'}) may only be solved by $\Gamma {\bf M}^{\dagger^{-1}}=j\beta\overline{T}_{4}$, per Lemma \ref{lemma1}, we have that $\Gamma=(j\beta\overline{T}_{4}){\bf M}^{\dagger}$ must be satisfied.
\end{corollary}

\subsection{Quadratic Energy Functions in a Classical Power System}
We now assume the classical model of a lossy multimachine power system model~\cite{Anderson:2008} and allow for constant power loads to be present. The forms of the FRFs associated with constant power loads (${\mathcal Y}_p$), constant impedance lines/shunts (${\mathcal Y}_z$), and classical generators (${\mathcal Y}_g$) are given in~\cite{Chevalier:2019P}. The set of plausible FRFs associated with these three elements may be constructed according to the following basis matrix combinations:
\begin{subequations}
\begin{align}
\mathcal{Y}_{p} & =\sum_{i=2,\,3}a_{i}\overline{T}_{i},\;\;a_i\!\in\! {\mathbb R}^1\label{eq: const_power}\\
\mathcal{Y}_{z} & =\sum_{i=1,\,4}a_{i}\overline{T}_{i},\;\;a_i\!\in\! {\mathbb R}^1\label{eq: const_impedance}\\
\mathcal{Y}_{g} & =\sum_{i=2,\,3,\,4}\left(a_{i}+jb_{i}\right)\overline{T}_{i},\;\;a_i,b_i\!\in\! {\mathbb R}^1\label{eq: Yg_sum}.
\end{align}
\end{subequations}
The FRF of a classical generator is derived and fully explained in~\cite{Chevalier:2018}, and, for convenience, is explicitly re-stated here:
\begin{align}
\mathcal{Y}_g \label{eq: Yg} =&\gamma(\Omega)\!\!\underbrace{\left[\!\!\begin{array}{cc}
\sin(\delta)\cos(\delta) \!\!\!&\!\!\! -\cos^{2}(\delta)\\
\sin^{2}(\delta) \!\!\!&\!\!\! -\sin(\delta)\cos(\delta)
\end{array}\!\!\right]}_{T_{\delta}}\!+\!\underbrace{\left[\!\!\begin{array}{cc}
0 \!\!&\!\! \frac{1}{X_{d}'}\\
\frac{-1}{X_{d}'} \!\!&\!\! 0
\end{array}\!\!\right]}_{T_X}\\
\gamma(\Omega) &=\frac{{\rm E}'^{2}}{X_{d}'^{2}}\frac{\left(M\left(j\Omega\right)^{2}+\frac{{\rm V}_{t}{\rm E}'}{X_{d}'}\cos(\varphi)\right)-j\left(\Omega D\right)}{\left(\frac{{\rm V}_{t}{\rm E}'}{X_{d}'}\cos(\varphi)-M\Omega^{2}\right)^{2}+\left(\Omega D\right)^{2}}.
\end{align}
where $\delta$ is the generator's absolute rotor angle and $\gamma\equiv\gamma(\Omega)\!\in\!{\mathbb C}^1$ is a complex frequency dependent parameter. Matrix ($\ref{eq: Yg}$) has many useful properties, one of which will be addressed in subsection \ref{Inf_Analysis}.

Given that \textit{any} admittance matrix $\mathcal Y$ may be written as the complex sum of the four weighted basis matrices, we introduce the following useful definition:
\begin{definition}\label{Def_power}
 We assume {some admittance $\mathcal Y$ may be written as} ${\mathcal Y}=\sum_{i=1}^{4}\left(a_{i}+jb_{i}\right)\overline{T}_{i}$. Using matrices $\bf M$ and $\Gamma$ from Corollary \ref{col_M_G}, where $\Gamma=(j\beta\overline{T}_{4}){\bf M}^{\dagger}$, we define $P^{\star}={\rm Re}\{{{\bf u}}^{\dagger}\left({\bf M}\mathcal{Y}\Gamma\right){{\bf u}}\}$ as the \textbf{dissipating power} for input vector ${\bf u}$. We further define two other types of quadratic power: \textbf{resistive power} $P_r^{\star}$ and \textbf{damping power} $P_d^{\star}$, where $P^{\star}=P_{r}^{\star}+P_{d}^{\star}$, and
\begin{itemize}
\item $P_{r}^{\star} ={\rm Re}\left\{ {{\bf u}}^{\dagger}\left({\bf M}\left(a_{1}\overline{T}_{1}\right)\Gamma\right){{\bf u}}\right\}$
\item $P_{d}^{\star} ={\rm Re}\left\{ {{\bf u}}^{\dagger}\left({\bf M}\left(jb_{2}\overline{T}_{2}+jb_{3}\overline{T}_{3}+jb_{4}\overline{T}_{4}\right)\Gamma\right){{\bf u}}\right\}$.
\end{itemize}
\end{definition}
We now prove that a perturbative system model containing elements (\ref{eq: const_power})-(\ref{eq: Yg_sum}) cannot be rendered passive under any quadratic passivity transformation. In other words, there is no quadratic quantity that is dissipated by all elements present in common networks. 

\begin{theorem}\label{theorem2}
There exist no non-singular matrices ${\bf M}\in\mathbb{C}^{2\times2}$
and $\Gamma\in\mathbb{C}^{2\times2}$ for which
\begin{align}
{\bf M}\mathcal{Y}\Gamma+\left({\bf M}\mathcal{Y}\Gamma\right)^{\dagger}\succeq0,\quad\forall\mathcal{Y}\in\left\{ \mathcal{Y}_{p},\;\mathcal{Y}_{z},\;\mathcal{Y}_{g}\right\}.
\end{align}
\begin{proof}
The FRF of a strictly reactive element, such as matrix $T_X$ in (\ref{eq: Yg}), is $\propto\overline{T}_{4}$ while the FRF of a strictly capacitive element is $\propto-\overline{T}_{4}$. The only way for ${\bf M}\overline{T}_{4}\Gamma+\left({\bf M}\overline{T}_{4}\Gamma\right)^{\dagger}\succeq0$ and $-{\bf M}\overline{T}_{4}\Gamma-\left({\bf M}\overline{T}_{4}\Gamma\right)^{\dagger}\succeq0$ to be simultaneously true is for ${\bf M}\overline{T}_{4}\Gamma+\left({\bf M}\overline{T}_{4}\Gamma\right)^{\dagger}\equiv{\bf 0}$. Since both reactive and capacitive elements appear in classical power systems, $\overline{T}_{4}$ must be lossless under the desired passivity transformation.

We now consider some classical generator whose damping characteristics are sufficiently small ($D\approx0$), such that $\gamma$ is a real parameter. In this case, the matrix ${\bf M}\mathcal{Y}_{g}\Gamma+\left({\bf M}\mathcal{Y}_{g}\Gamma\right)^{\dagger}$ reduces to ${\bf M} \gamma T_\delta \Gamma+\left({\bf M} \gamma T_\delta \Gamma\right)^{\dagger}$ since ${T_X}$ must be a lossless element according to the previous conclusion about $\overline{T}_4$.

We define the squared electromechanical resonant frequency associated with the classical generator as $\Omega^2_{r}=\frac{{\rm V}{\rm E}'}{MX'_{d}}\cos(\varphi)$. For some $\epsilon$, $\gamma(\Omega_{r}-\epsilon)=-\gamma(\Omega_{r}+\epsilon)$. We must therefore ensure that ${\bf M} \gamma T_\delta \Gamma+\left({\bf M} \gamma T_\delta \Gamma\right)^{\dagger}\succeq0$ and $-{\bf M} \gamma T_\delta \Gamma-\left({\bf M} \gamma T_\delta \Gamma\right)^{\dagger}\succeq0$, respectively, when ensuring the generator's passivity on either side of the resonant peak. The only way for these statements to be simultaneously true is for ${\bf M} T_\delta \Gamma+\left({\bf M} T_\delta \Gamma\right)^{\dagger}\equiv{\bf0}$. To accomplish this, we consider the numerical structure of $T_{\delta}$ for two plausible rotor angle values: $\delta_1=0$ and $\delta_2=\frac{\pi}{4}$:
\begin{align}
\label{eq: del1}T_{\delta}\left(\delta_{1}\right) & =\left[\begin{array}{cc}
0 & -1\\
0 & 0
\end{array}\right]=\frac{1}{2}\left(\overline{T}_{4}-\overline{T}_{3}\right),\\
\label{eq: del2}T_{\delta}\left(\delta_{2}\right) & =\left[\begin{array}{cc}
\frac{1}{2} & -\frac{1}{2}\\
\frac{1}{2} & -\frac{1}{2}
\end{array}\right]=\frac{1}{4}\left(\overline{T}_{2}+\overline{T}_{4}\right).
\end{align}
Since ${\bf M} \overline{T}_{4} \Gamma+\left({\bf M} \overline{T}_{4} \Gamma\right)^{\dagger}\equiv{\bf0}$, then we must also require ${\bf M} \overline{T}_{3} \Gamma+\left({\bf M} \overline{T}_{3} \Gamma\right)^{\dagger}\equiv{\bf0}$ from (\ref{eq: del1}) and ${\bf M} \overline{T}_{2} \Gamma+\left({\bf M} \overline{T}_{2} \Gamma\right)^{\dagger}\equiv{\bf0}$ from (\ref{eq: del2}) in order to ensure that ${\bf M} T_{\delta} \Gamma+\left({\bf M} T_{\delta} \Gamma\right)^{\dagger}\equiv{\bf0}$. We are thus requiring that ${\rm Re}\left\{ \tilde{{\bf v}}^{\dagger}\left({\bf M}\overline{T}_{i}\Gamma\right)\tilde{{\bf v}}\right\}  =0,\;\forall\tilde{{\bf v}}\in\mathbb{C}^{2\times1},\,\forall i\in\{2,3,4\}$. As stated in Corollary \ref{col_M_G}, the only way to achieve losslessness for basis matrices 2, 3 and 4 is for $\Gamma=(j\beta\overline{T}_{4}){\bf M}^{\dagger}$. By employing this transformation, Lemma \ref{lemma1} proves that the quadratic energy associated with any element containing $\overline{T}_{1}$ will be rendered indefinite in sign. Since (\ref{eq: const_impedance}) contains $\overline{T}_{1}$ when resistance is present in the network, then there exists no nonsingular matrices ${\bf M}$ and $\Gamma$ for which ${\bf M}\mathcal{Y}\Gamma+\left({\bf M}\mathcal{Y}\Gamma\right)^{\dagger}\succeq0,\quad\forall\mathcal{Y}\in\left\{ \mathcal{Y}_{p},\;\mathcal{Y}_{z},\;\mathcal{Y}_{g}\right\}$.
\end{proof}
\end{theorem}

\begin{corollary}
By choosing ${\bf M}=\overline{T}_1$, $\beta=\frac{-1}{\Omega}$, and $\Gamma=(j\beta\overline{T}_{4}){\bf M}^{\dagger}$, we arrive at the passivity transformation implicitly employed by the DEF, as given by\cite[eq. (38a,b)]{Chevalier:2019P}.
\end{corollary}

\begin{corollary}\label{corollary_PES}
The passivity transformation $\Gamma=(j\beta\overline{T}_{4}){\bf M}^{\dagger}$ renders constant power and reactive (capacitive or inductive) elements of the system lossless, classical generators passive (see\cite[eq. (49)]{Chevalier:2019P}), and resistive elements non-passive. Without resistive elements, the DEF method is a fully reliable source location technique in classical power systems.
\end{corollary}

\begin{corollary}\label{corollary_ZIP}
Since the linearized admittance matrix ${\mathcal Y}$ associated with any ZIP load is purely real, i.e. ${\mathcal Y}={\rm Re}\{{\mathcal Y}\}$, the eigenvalues of its transformed Hermitian part will be equal and opposite in value: $\lambda({\bf M}\mathcal{Y}\Gamma+\left({\bf M}\mathcal{Y}\Gamma\right)^{\dagger})=\pm\alpha$, {where $\alpha=0$ for constant power or purely reactive loads.}
\end{corollary}

If a network has no resistive elements and is truly passive, no quadratic energy production can occur on regular network elements, so injections of energy have to be related to external sources, like FOs. This is why finding a passive energy-like form is important, and why the almost-passive form used by DEF has had so much success. To further show why the results of Theorem \ref{theorem2} are problematic for the DEF method, we consider the structure of (\ref{eq: DEF_f}): since the generator damping contributions are positive definite, the source energy is necessarily negative in a power system {with no resistance}. {This is not true in a lossy power system. To show why not,} we define block matrices $\underline{\bf M}\in{\mathbb C}^{2n\times2n}$ and $\underline{\Gamma}\in{\mathbb C}^{2n\times2n}$, where
\begin{align*}
\begin{array}{cc}
\underline{\bf M}=\left[\begin{array}{ccc}
{\bf M}\!\!\! &  & \!\!\!{\bf 0}\vspace{-0.15cm}\\
 & \ddots\vspace{-0.15cm}\\
{\bf 0}\!\!\! &  & \!\!\!{\bf M}
\end{array}\right],\;\; & \underline{\Gamma}=\left[\begin{array}{ccc}
\Gamma\!\!\! &  & \!\!\!{\bf 0}\vspace{-0.15cm}\\
 & \ddots\vspace{-0.15cm}\\
{\bf 0}\!\!\! &  & \!\!\!\Gamma
\end{array}\right],\end{array}
\end{align*}
and whose block diagonal matrices are given by ${\bf M}$ and $\Gamma=(j\beta\overline{T}_{4}){\bf M}^{\dagger}$, respectively. We left multiply (\ref{eq: J=YV'}) by $\underline{{\bf M}}$ and insert $\underline{\Gamma}\underline{\Gamma}^{-1}$ on the RHS:
\begin{align}\label{eq: MJ'}
-\underline{{\bf M}}{\bf J}'=\underline{{\bf M}}\left(E_{a}^{\dagger}\mathcal{Y}_{L}E_{a}+\mathcal{Y}_{S}'\right)\underline{\Gamma}\underline{\Gamma}^{-1}{\bf V}_{b}.
\end{align}
Defining the transformed voltage vectors ${\bf U}_b=\underline{\Gamma}^{-1}{\bf V}_{b}$ and ${\bf U}_s={\Gamma}^{-1}{\bf V}_{s}$, we left multiply (\ref{eq: MJ'}) by ${\bf U}_b^{\dagger}$ and simplify. We may group the dissipating power injections into their respective contributing groups (assuming lossless loads):
\begin{align}
0\!=\!\underbrace{{\bf U}_{s}^{\dagger}{\bf M}{\bm{\mathcal I}}'}_\text{Source}\!+\!\underbrace{{\bf U}_{b}^{\dagger}(\underline{{\bf M}}\mathcal{Y}_{S}'\underline{\Gamma}){\bf U}_{b}}_\text{Generator}\!+\!\underbrace{{\bf U}_{b}^{\dagger}(\underline{{\bf M}}E_{a}^{\dagger}\mathcal{Y}_{L}E_{a}\underline{\Gamma}){\bf U}_{b}}_\text{Network}.\label{eq: SGN}
\end{align}
The FO source term can produce only negative damping energy, i.e. ${\rm Re}\{{\bf U}_{s}^{\dagger}{\bf M}{\bm{\mathcal I}}'\}{<}0$, if the condition
\begin{align}\label{eq: damping_condition}
\!\!\underline{{\bf M}}\left(\mathcal{Y}_{S}'\!+\!E_{a}^{\dagger}\mathcal{Y}_{L}E_{a}\right)\underline{\Gamma}+\left(\underline{{\bf M}}\left(\mathcal{Y}_{S}'\!+\!E_{a}^{\dagger}\mathcal{Y}_{L}E_{a}\right)\underline{\Gamma}\right)^{\dagger}\!\succ0
\end{align}
is met. If it is not met, indefinitely signed resistive energy can dominate damper winding energy absorption and the source term can, in fact, appear as a positively damped element. In this plausible situation, the DEF method will fail. {We note that violation of (\ref{eq: damping_condition}) is a necessary but not sufficient condition for DEF failure.} Next, we show that the sign of the injected resistive energy {associated with a transmission network} can be negated if all system voltages are complex conjugated.

\begin{theorem}\label{theorem_U*}
Consider a lossy transmission {network (just the network)}. For any transformed voltage vector ${\bf U}_{b}$ which yields quadratic energy ${\rm Re}\{{\bf U}_{b}^{\dagger}(\underline{{\bf M}}(E_{a}^{\dagger}\mathcal{Y}_{L}E_{a})\underline{\Gamma}){\bf U}_{b}\} \!=\! P_r^{\star}$, there exists conjugated vector ${\bf U}^*_{b}$ which yields an equal and opposite quadratic energy ${\rm Re}\{{\bf U}_{b}^{*\dagger}(\underline{{\bf M}}(E_{a}^{\dagger}\mathcal{Y}_{L}E_{a})\underline{\Gamma}){\bf U}_{b}^{*}\}=-P_r^{\star}$, where $\underline{{\bf M}}$ and $\underline{\Gamma}$, with submatrices $\bf M$ and $\Gamma$, yield from Corollary \ref{col_M_G}.
\begin{proof}
We split the transmission line matrix into its conductive and susceptive parts: $\mathcal{Y}_{L}=\mathcal{Y}_{\bar G}+\mathcal{Y}_{\bar B}$, where $\mathcal{Y}_{{\bar G},i}=G_i\overline{T}_1$ and $\mathcal{Y}_{{\bar B},i}=B_i\overline{T}_4$. We also define block matrices ${\bf M}=j\overline{T}_4\Gamma$ and $\Gamma=\overline{T}_1$ from $\underline{\bf M}$ and $\underline{\Gamma}$. Therefore, $\underline{{\bf M}}\left(E_{a}^{\dagger}\mathcal{Y}_{L}E_{a}\right)\underline{\Gamma} =\underline{{\bf M}}\left(E_{a}^{\dagger}\left(\mathcal{Y}_{\bar G}+\mathcal{Y}_{\bar B}\right)E_{a}\right)$. The Hermitian part (termed ${\bf H}$) is
\begin{subequations}
\begin{align}
{\bf H} & =\underline{{\bf M}}E_{a}^{\dagger}({\scriptstyle \frac{\mathcal{Y}_{G}+\mathcal{Y}_{Bx}}{2}})E_{a}+E_{a}^{\dagger}({\scriptstyle \frac{\mathcal{Y}_{G}-\mathcal{Y}_{Bx}}{2}})E_{a}\underline{{\bf M}}^{\dagger}\\
 & =\underline{{\bf M}}\left(E_{a}^{\dagger}{\scriptstyle \frac{\mathcal{Y}_{G}}{2}}E_{a}\right)+\left(E_{a}^{\dagger}{\scriptstyle \frac{\mathcal{Y}_{G}}{2}}E_{a}\right)\underline{{\bf M}}\label{eq: Herm_Comm}\\
 & =\underline{{\bf M}}\left(E_{a}^{\dagger}\mathcal{Y}_{G}E_{a}\right)\label{eq: Herm_Comm2}
\end{align}
\end{subequations}
where the matrices of (\ref{eq: Herm_Comm}) commute since the product of Hermitian matrices is also Hermitian. We define ${\overline G}_{ij}\equiv j G_{ij}{\overline T}_4$, where $G_{ij}$ is the scalar line conductance connecting buses $i$ and $j$, and ${\bf u}_{i}\!\subset\!{\bf U}_b$ is the voltage element of ${\bf U}_b$ associated with bus $i$. From (\ref{eq: Herm_Comm2}), the quadratic power is
\begin{subequations}
\begin{align}
P_r^{\star} \!& =\!\sum_{i=1}^{n}{\bf u}_{i}^{\dagger}(\sum_{j\ne i}\overline{G}_{ij}){\bf u}_{i}-\sum_{j\ne i}^{n}\sum_{i\ne j}^{n}{\bf u}_{i}^{\dagger}\overline{T}_{4}\overline{G}_{ij}{\bf u}_{j}\\
 & =\!\sum_{i,j\in\mathcal{E}}({\bf u}_{i}^{\dagger}\overline{G}_{ij}{\bf u}_{i}\!+\!{\bf u}_{j}^{\dagger}\overline{G}_{ij}{\bf u}_{j}\!-\!{\bf u}_{i}^{\dagger}\overline{G}_{ij}{\bf u}_{j}\!-\!{\bf u}_{j}^{\dagger}\overline{G}_{ij}{\bf u}_{i})\\
 & =\!\sum_{i,j\in\mathcal{E}}({\bf u}_{i}-{\bf u}_{j})^{\dagger}\overline{G}_{ij}\underbrace{({\bf u}_{i}-{\bf u}_{j})}_{{\bf u}_{ij}}.\label{eq: Pr_conj}
\end{align}
\end{subequations}
The quadratic quantity ${\bf x}^{\dagger} {\overline G}_{ij}{\bf x}=\epsilon$ may be negated by conjugating the input (proof trivial): ${\bf x}^{*\dagger} {\overline G}_{ij}{\bf x}^*\!=\!-\epsilon$. By taking ${\bf U}_b^*$ as an input to ${\rm Re}\{\underline{\bf M}{\bf U}_{b}^{*\dagger}(E_{a}^{\dagger}\mathcal{Y}_{G}E_{a}){\bf U}_{b}^{*}\}$, by (\ref{eq: Pr_conj}) we {thus have $\sum_{i,j\in\mathcal{E}}{\bf u}_{ij}^{*\dagger}\overline{G}_{ij}{\bf u}_{ij}^{*} = -P_{r}^{\star}$}.
\end{proof}
\end{theorem}

We note that ${\bf U}_b$ cannot be chosen arbitrarily; it must represent a valid solution to the linear system of (\ref{eq: J=YV'}). Since ${\bf U}_b$ is not itself a degree of freedom but rather a response to some current injection, statements about the mathematical characteristics of ${\bf M}\mathcal{Y}_{N}\Gamma + ({\bf M}\mathcal{Y}_{N}\Gamma)^{\dagger}$ are difficult to prove using energy-based arguments. {Despite this fact, the following two theorems offer useful insights into these characteristics. First, we offer a definition of an associated system.}
\begin{definition}
{Consider classical power system $\Sigma_c$ with ZIP loads, passive generators, and a lossy transmission network. This system's voltage vector ${\bf V}_b$ is ${\bf V}_b={\bf V}_{s}^{\frown}{\bf V}_{ns}$, where ${\bf V}_{ns}=-{\mathcal Y}_4^{-1}{\mathcal Y}_3{\bf V}_s$ from (\ref{eq: Net_Demcop'}). All voltage vectors are transformed such that ${\bf U} = \Gamma^{-1}{\bf V}$. The Kron reduced admittance (DWE) seen by a FO source is ${\mathcal Y}_N$ as in (\ref{eq: YN1}). The matrix ${\bf N}_c=\frac{1}{2}({\bf M}\mathcal{Y}_{N}\Gamma)+\frac{1}{2}({\bf M}\mathcal{Y}_{N}\Gamma)^{\dagger}$, with matrices $\bf M$ and $\Gamma$ from Corollary \ref{col_M_G}, will have two eigenvalues: $\lambda_1$ and $\lambda_2$.}
\end{definition}

\begin{theorem}\label{theorem3}
{Consider ${\bf N}_c$ from $\Sigma_c$}:
\begin{itemize}
\item[$(a)$] If condition (\ref{eq: damping_condition}) is met, $\lambda_1$, $\lambda_2 > 0$.
\item[$(b)$] If ${\rm Re}\{{\bf U}_{s}^{\dagger}{\bf M}{\bm{\mathcal I}}'\}>0$ is measured, i.e. the DEF method has failed, condition (\ref{eq: damping_condition}) is violated and $\lambda_1 \lor \lambda_2 <0$.
\end{itemize}
\begin{proof}
By modifying (\ref{eq: SGN}), we have $-{\bf U}_{s}^{\dagger}{\bf M}{\bm{\mathcal I}}'\!=\!{\bf U}_{b}^{\dagger}\underline{{\bf M}}(\mathcal{Y}_{S}'\!+\!E_{a}^{\dagger}\mathcal{Y}_{L}E_{a})\underline{\Gamma}{\bf U}_{b}$, and with condition (\ref{eq: damping_condition}) met, $-{\bf U}_{s}^{\dagger}{\bf M}{\bm{\mathcal I}}'>0$. Since $-{\bm{\mathcal I}}'={\mathcal Y}_N {\bf V}_s$, we have that ${\bf U}_{s}^{\dagger}{\bf M}\mathcal{Y}_{N}\Gamma{\bf U}_{s} >0$ by substitution, implying that $\lambda\{({\bf M}\mathcal{Y}_{N}\Gamma)+({\bf M}\mathcal{Y}_{N}\Gamma)^{\dagger}\}>0, \;\forall \lambda\in\{\lambda_1,\lambda_2\}$, proving proposition $(a)$.

If ${\rm Re}\{{\bf U}_{s}^{\dagger}{\bf M}{\bm{\mathcal I}}'\}>0$, then ${\bf U}_{s}^{\dagger}{{\bf N}_c}{\bf U}_{s}{<}0$ is directly implied. Accordingly, $\lambda_1 \lor \lambda_2 <0$. By the conservation argument presented in the proof of proposition $(a)$, if ${\bf U}_{s}^{\dagger}{{\bf N}_c}{\bf U}_{s}{<}0$, then ${\rm Re}\{{\bf U}_{b}^{\dagger}\underline{{\bf M}}(\mathcal{Y}_{S}'\!+\!E_{a}^{\dagger}\mathcal{Y}_{L}E_{a})\underline{\Gamma}{\bf U}_{b}\}<0$ implying the violation of condition (\ref{eq: damping_condition}), proving proposition $(b)$.
\end{proof}
\end{theorem}

{Our final theorem characterizes how the eigenvalues change as resistance is added to $\Sigma_c$.}

\begin{theorem}\label{theorem4}
{Consider ${\bf N}_c$ from $\Sigma_c$. No amount of additional resistance to lines or loads can cause the eigenvalues of ${\bf N}_c$ to both become negative: if ${\rm det}({\bf N}_c)\ge0$ then ${\rm trace}({\bf N}_c)\ge0$.}
\begin{proof}
{Consider some altered version of $\Sigma_c$ where all resistance has been removed from the system. In this situation, $\lambda_1,\lambda_2 \ge 0$ according to Theorem \ref{theorem3}.\\
\indent Consider some secondary altered version of $\Sigma_c$ where generators have no damping (making them ``lossless"). In this situation, matrix ${\mathcal Y}_N$ will be purely real (${\mathcal Y}_N\!\in\!{\mathbb R}^{2\times2}$) via (\ref{eq: YN1}). Therefore, assuming lossy lines, $\lambda_1=-\lambda_2\ne0$.\\
\indent All real systems exist between these two alternatives. We now consider a system with nominal resistance and some value $\alpha$ which parameterizes the amount of damping in the system generators ($\alpha=0$ corresponds to no damping). When $\alpha=0$, $\lambda_1=-\lambda_2\ne0$. As $\alpha\rightarrow\infty$, then $\lambda_1,\,\lambda_2\rightarrow{\mathbb R}^+$. In between these extremes, if for some value of $\alpha > 0$, there is $\lambda_1,\,\lambda_2\in{\mathbb R}^-$, then it would imply that the addition of positive damping causes the system to lose the ability to dissipate the quadratic energy which the generator damping consumes. This is a contradiction, so at least one eigenvalue must always remain positive for all levels of damping and resistance.}
\end{proof}
\end{theorem}
\begin{remark}\label{Remark: never_passive}
    Since the DWE ${\mathcal Y}_N$ is the admittance of the network ``seen" by the source, the negative DWE $-{\mathcal Y}_N$ is the admittance ``seen" by the network behind the source bus. {By Theorem \ref{theorem4}, the eigenvalues of ${\bf N}_c=\frac{1}{2}({\bf M}\mathcal{Y}_{N}\Gamma)+\frac{1}{2}({\bf M}\mathcal{Y}_{N}\Gamma)^{\dagger}$ can never be simultaneously negative. Stated differently, the eigenvalues of $-{\bf N}_c$ can never be simultaneously positive. \textbf{For this reason, the admittance seen by the network behind the FO source can never be truly passive.}}
\end{remark}

{\subsection{Inference Analysis Observations}\label{Inf_Analysis}
If a system-wide admittance matrix model, i.e. (\ref{eq: Yb_Model}), is fully known at the time of a FO event, locating the source of the FO is trivial once PMU measurements are obtained. Such a model is often unknown or poorly known in real time, so model based approaches aren't typically applicable. Using PMU inference methods to build the admittance model (in real time) are thus a tempting solution. Unfortunately, the system is inherently underdetermined. In a standard linear circuit, admittance may be computed as the simple ratio of complex inputs and complex outputs. This cannot be done in a MIMO system.} Assuming complex input (${\tilde{\bf u}}_v$) and output (${\tilde{\bf y}}$) vectors for the system in (\ref{eq: Y(jOmega)}) are given, a generalized admittance ${\mathcal Y}\in{\mathbb C}^{2\times2}$ cannot be directly inferred since (\ref{eq: Y(jOmega)}) represents four real linear equations while ${\mathcal Y}$ is specified with 8 coefficients ($a_{i},\,b_{i},\, i\in\{1,2,3,4\}$):
\begin{align}\label{eq: Y_inf}
\left[\begin{array}{cc}
\mathcal{Y}_{1} & \mathcal{Y}_{2}\\
\mathcal{Y}_{3} & \mathcal{Y}_{4}
\end{array}\right]=\sum_{i=1}^{4}\left(a_{i}+jb_{i}\right)\overline{T}_{i}.
\end{align}
The system is thus severely underdetermined. {For a classical generator, though, this is not the case. This is an important observation, because the dynamics of a system experiencing a FO are dominated by the electromechanical response of synchronous generators.
\begin{proposition}\label{prop: Yg}
The admittance ${\mathcal Y}_g$ of a classical generator can be uniquely specified as the weighted sum of six basis matrices with only \textbf{four} real coefficients.
\begin{proof}
It may be observed that ${\mathcal Y}_g$ in ($\ref{eq: Yg}$) contains no complex combination of basis matrix $\overline{T}_{1}$. Therefore, ${\mathcal Y}_g$ may be written using six basis matrix coefficients $a_{i}$, $b_{i}$, $i\in\{2,3,4\}$ via $\mathcal{Y}_{g}=\sum_{i=2...4}\left(a_{i}+jb_{i}\right)\overline{T}_{i}.$ We further observe that for $(a_{2} + jb_{2})\overline{T}_{2}$ and $(a_{3} + jb_{3})\overline{T}_{3}$, the ratios of $a_{2}/b_{2}$ and $a_{3}/b_{3}$ must be equal, so $a_{2}=b_{2}a_{3}/b_{3}$. This eliminates one coefficient. Finally, we write $\gamma = \gamma_r+j\gamma_i$ and observe that $b_2=\gamma_i\sin(\delta)\cos(\delta)$ while $b_3+b_4=\gamma_i\sin^2(\delta)$ and $b_3-b_4=-\gamma_i\cos^2(\delta)$. Solving this system yields $b_4=+\sqrt{b_2^2+b_3^2}$ (assuming damping $D$ is positive). Therefore, ${\mathcal Y}_g$ may be specified with just four coefficients: $a_{3}$, $a_{4}$, $b_{2}$, and $b_{3}$.
\end{proof}
\end{proposition}
\begin{corollary}
From Proposition \ref{prop: Yg}, classical generator admittance ${\mathcal Y}_g$ can be specified with only four parameters. Thus, the inference problem $\min_{{\mathcal Y}_g}\{{\tilde{\bf y}} - {\mathcal Y}_g{\tilde{\bf u}}_v\}$ represents 4 equations and 4 unknowns and is thus a ``determined" problem.
\end{corollary}
In other words, the admittance of a classical generator can be fully inferred from PMU data.}

\section{{Practical Applications of Passivity Analysis}}\label{Practical Applications}
{In this section, we expound three practical applications of the analysis presented in this paper. First, we explicitly state how to implement the DEF source location method via the proposed frequency domain methods. Second, we show how system operators can use the proposed framework to predict how the DEF method will perform in their respective networks without performing simulations. And third, we explain how the tools in this paper can be used to understand how the addition of new grid components will affect performance of the DEF method. In all further analysis, we assume 
\begin{align}
    {\bf M}&=\left[\!\begin{array}{cc}
    0 & j\\
    -j & 0
    \end{array}\!\right].
\end{align}}

\subsection{{Energy Flow Calculations in the Frequency Domain}}
{In order to compute the dissipating power signal $P^{\star}$ in a system which is experiencing a FO event, we convert voltage and current PMU data into rectangular coordinates. Next, we take the Fast Fourier Transform (FFT) of these signals, thus constructing ${\tilde {\bf V}}(\Omega)=[{\tilde V}_r(\Omega),\,{\tilde V}_i(\Omega)]^{\top}$ and ${\tilde {\bf I}}(\Omega)=[{\tilde I}_r(\Omega),\,{\tilde I}_i(\Omega)]^{\top}$ at all relevant buses and lines. Finally, we evaluate these signals at the forcing frequency $\Omega_d$ in order to compute the dissipating power injection at each bus or flow along each line:
\begin{align}\label{eq: DEF_passivity}
    P^{\star}&={\rm Re}\{\tilde{{\bf V}}(\Omega_{d})^{\dagger}{\bf M}\tilde{{\bf I}}(\Omega_{d})\}.
\end{align}
Positive $P^{\star}$ flowing out of an element indicates the presence of a non-passive component (typically a FO source). We refer readers to Appendix \ref{AppB} for a discussion on the difficulties of dealing with rectangular coordinates in a realistic power system. Using this procedure, a map, such as the one presented in Fig. \ref{fig: Full_Sys_nR}, can be constructed.}

\subsection{Predicting Performance of DEF Source Location}
The implicit goal of the DEF method is to locate the source of negative damping in the system. When resistive elements are introduced to the network, the goal becomes obfuscated because the source {may appear} passive, and thus, positively damped. By {Remark \ref{Remark: never_passive}} though, the negative DWE of the source cannot be \textit{physically} passive. A key contribution of this paper recognizes that while generators continue to be physically passive, the source only appears\footnote{The transformed FRF will have one positive and one negative eigenvalue.} passive when the DEF method fails in a classical power system.

{Before system operators can consistently rely on using the DEF method for FO source location, it is important that they first test if the method will perform adequately in their respective networks. Algorithm \ref{alg: DEF_Check} clarifies the exact analytical procedure which can be used to test if the DEF method will perform successfully for any given bus in the network at any given FO frequency. We note that this off-line analysis in not restricted to classical power systems: any system component can be incorporated so long as its dynamics can be analytically approximated and linearized. Additionally, the \textit{type} of oscillation source is entirely arbitrary: the methods in Algorithm \ref{alg: DEF_Check} predict the sign of the energy flowing from the source bus terminal as a function of how the system dynamically responds to the abstracted oscillation signal. It is thus similar to testing if condition (\ref{eq: damping_condition}) is violated.}
\begin{algorithm}\label{alg: DEF_Check}
\caption{{Reliability Test for Energy Flow Methods}}
{\textbf{START}\\
\begin{enumerate}[label=\textbf{\arabic*},start=1]
    \item construct nonlinear model (\ref{eq: nonlin_sys}) for all shunt\\ components and linearize to build admittance (\ref{eq: Y(jOmega)})
    \item construct shunt admittance matrix (\ref{eq: YS_mat}), augmented incidence matrix (\ref{eq: Ea}) and line admittance matrix (\ref{eq: YL_mat})
    \item construct dynamic nodal admittance (\ref{eq: Yb_Model})
\end{enumerate}
\For{each plausible FO frequency $\Omega_d$}{
\For{each plausible source bus $i$}{
\begin{enumerate}
\item remove shunt admittance ${\mathcal Y}_{s,i}$ from ${\mathcal Y}_{S}$
\item partition network via (\ref{eq: Net_Demcop_both}) and perform\\ Kron reduction via (\ref{eq: YN1}) to construct ${\mathcal Y}_N$
\item compute $\lambda_{1,2}=\lambda\{{\bf M}{\mathcal Y}_N + ({\bf M}{\mathcal Y}_N)^\dagger\}$
\end{enumerate}
\textbf{Assume bus} $i$ \textbf{generates FO of frequency} $\Omega_d$:\\
  \uIf{$\lambda_1,\lambda_2\ge0$}{DEF will succeed}
  \uElseIf{$\lambda_1,\lambda_2\le0$}{DEF will be fail}
  \Else{DEF will perform unreliably}}}}
\end{algorithm}

\subsection{{Analysis of Additional Grid Components}}
{Until this point in the paper, only the elements of a lossy classical power system have been considered, but it is important for system operators to consider how the addition of new elements will effect energy-based source location methods. If newly added elements are non-passive, it is also important for system operators to test the degree of allowed penetration from these elements before energy-based source location methods will fail. In this subsection, we present examples of passivity analysis applied to three non-classical elements: frequency dependent loads, droop-controlled inverter systems, and third order synchronous generators with first order Automatic Voltage Regulators (AVRs).}

\subsubsection{{Frequency Dependent Load}}
Assuming power is an instantaneous function of voltage magnitude and frequency\cite{Milano:2013}, we may write $P(t)$ and $Q(t)$ via
\begin{subequations}\label{eq: PQ}
\begin{align}\label{eq: P(t)}
P(t) & =P_{0}\left({\rm V}/{\rm V}_{0}\right)^{\alpha_{p}}\left(\omega/\omega_{0}\right)^{\beta_{p}}\\
Q(t) & =Q_{0}\left({\rm V}/{\rm V}_{0}\right)^{\alpha_{q}}\left(\omega/\omega_{0}\right)^{\beta_{q}},\label{eq: Q(t)}
\end{align}
\end{subequations}
where $\omega=\omega_0+\dot\theta$. Since we are interested in the linearized responses of (\ref{eq: P(t)}) and (\ref{eq: Q(t)}), we evaluate their partial derivatives at equilibrium (${\rm V}_0$, $\omega_0{=1}$).
Assuming sinusoidal perturbations, phasor notation yields
\begin{align}\label{eq: Ya}
\underbrace{\left[\begin{array}{c}
\tilde{P}\\
\tilde{Q}
\end{array}\right]}_{\tilde{{\bf S}}} & =\underbrace{\left[\begin{array}{cc}
{\frac{\alpha_{p}}{{\rm V}_0}}P_{0} & \beta_{p}P_{0}\\
{\frac{\alpha_{q}}{{\rm V}_0}}Q_{0} & \beta_{q}Q_{0}
\end{array}\right]}_{{\mathcal Y}_{a}}\underbrace{\left[\begin{array}{cc}
1 & 0\\
0 & j
\end{array}\right]}_{I_{j}}\underbrace{\left[\begin{array}{c}
\tilde{{\rm V}}\\
\tilde{\theta}
\end{array}\right]}_{\tilde{{\bf V}}_p}
\end{align}
where $\tilde{\omega}=j\tilde{\theta}$. Employing matrix ${\rm T}_1$ from {Appendix \ref{AppB}}, we transform from polar to rectangular via $\tilde{{\bf V}}={\rm T}_{1}\tilde{{\bf V}}_p$ in (\ref{eq: Ya}):
\begin{align}
\tilde{{\bf S}} & ={\mathcal Y}_{a}I_{j}{\rm T}_{1}^{-1}\tilde{{\bf V}}\label{eq: S_tilde}.
\end{align}
We next consider perturbations of $P=V_{r}I_{r}+I_{i}V_{i}$ and $Q=V_{i}I_{r}-V_{r}I_{i}$. Treating $\bar{V}_{i/r}$ and $\bar{I}_{i/r}$ as steady state values, we linearize and convert to phasor notation:
\begin{align}\label{eq: PQ_phasors}
\left[\!\begin{array}{c}
\tilde{P}\\
\tilde{Q}
\end{array}\!\right]=\underbrace{\left[\!\begin{array}{cc}
\bar{I}_{r} & \bar{I}_{i}\\
-\bar{I}_{i} & \bar{I}_{r}
\end{array}\!\right]}_{A_{I}}\underbrace{\left[\!\begin{array}{c}
\tilde{V}_{r}\\
\tilde{V}_{i}
\end{array}\!\right]}_{\tilde{\bf V}}+\underbrace{\left[\!\begin{array}{cc}
\bar{V}_{r} & \bar{V}_{i}\\
\bar{V}_{i} & -\bar{V}_{r}
\end{array}\!\right]}_{A_{V}}\underbrace{\left[\!\begin{array}{c}
\tilde{I}_{r}\\
\tilde{I}_{i}
\end{array}\!\right]}_{\tilde{\bf I}}.
\end{align}
By equating (\ref{eq: S_tilde}) and (\ref{eq: PQ_phasors}), ${\tilde {\bf I}}$ and ${\tilde {\bf V}}$ are directly related by
\begin{align}\label{eq: Yb} 
\tilde{{\bf I}}=\underbrace{A_{V}^{-1}\left({\mathcal Y}_{a}I_{j}{\rm T}_{1}^{-1}-A_{I}\right)}_{{\mathcal Y}_b}\tilde{{\bf V}}.
\end{align}
{Finally, we set $\alpha_p=\alpha_q=0$ to isolate the effects of frequency, and we compute eigenvalues $\lambda_{1,2}=\lambda\{{\bf M}\mathcal{Y}_{b}+({\bf M}\mathcal{Y}_{b})^{\dagger}\}$:
\begin{align}\label{eq: eigs_FD}
    \lambda_{1,2} &\!=\! \frac{\beta_{p}\cos(\varphi)\!\pm\!\sqrt{\!\frac{\beta_{p}^{2}}{2}\!\left(1\!+\!\cos(2\varphi)\!\right)\!+\!\frac{\beta_{q}^{2}}{2}\!\left(1\!\!-\!\cos(2\varphi)\!\right)}}{{\rm I}/{\rm V}}.
\end{align}
where $\varphi = \theta-\phi$. In setting $\beta\equiv\beta_{p}\approx\beta_{{q}}$, (\ref{eq: eigs_FD}) simplifies to
\begin{align}\label{eq: eigs_FDs}
   \lambda_{1,2} = \beta({\cos(\varphi)\pm1}){\rm V}/{\rm I}.
\end{align}
Since power factor is usually close to unity, this case of frequency dependent load is \textit{primarily passive} for $\beta>0$ because the positive eigenvalue will be far larger than the negative eigenvalue. Otherwise, if $\beta_p=0$, then the load behaves like a resistor (equal and opposite eigenvalues), but if instead $\beta_q=0$, then the load is truly passive for reasonable values of power factor.}

\subsubsection{{Droop-controlled Inverter}}
{We consider the dynamics of a droop-controlled inveter circuit; such circuits have the potential to dominate distributed energy resource interconnections. As given in~\cite{Vorobev:2019}, the circuit's impedance may be stated as 
\begin{align}\label{eq: Z_inverter}
    \mathcal{Z}=\left[\begin{array}{cc}
    R_{c}+j\Omega L_{c} & -X_{c}-\frac{k_{q}}{1+j\tau\Omega}\\
    X_{c}-\frac{k_{p}}{\tau\Omega^{2}-j\Omega} & R_{c}+j\Omega L_{c}\end{array}\right].
\end{align}
Associated parameters are explained in~\cite{Vorobev:2019}, but $k_p$ and $k_q$ are the active and reactive power droop coefficients, and $R_c$ is the virtual + coupling resistance. We seek to analyze the passivity of this system. Rather then invert (\ref{eq: Z_inverter}), we consider the passivity of the impedance $\mathcal Z$ directly. Appendix \ref{AppC} establishes the equivalence of passivity classifications between an admittance $\mathcal Y$ and its associated impedance ${\mathcal Z}={\mathcal Y}^{-1}$ for transformation matrix $\bf M$. We thus compute the Hermitian part of ${\bf M}\mathcal{Z}$:
\begin{align}\label{eq: GTI}
{\bf M}\mathcal{Z}+({\bf M}\mathcal{Z})^{\dagger}=2\left[\begin{array}{cc}
\frac{k_{p}}{\Omega((\tau\Omega)^{2}+1)} & jR_{c}\\
-jR_{c} & \frac{k_{q}\tau\Omega}{(\tau\Omega)^{2}+1}
\end{array}\right].
\end{align}
Since $\tau$, $\Omega$, and the droop coefficients are all positive, if $R_c$ is small, then the eigenvalues of (\ref{eq: GTI}) are simply
\begin{align}\label{eq: droop_evals}
    \lambda_{1,2}\approx k_{p}\frac{2}{\Omega((\tau\Omega)^{2}+1)},\,\,k_{q}\frac{2\tau\Omega}{(\tau\Omega)^{2}+1}
\end{align}
and the system is effectively passive. If $R_c$ is not neglected, though, the eigenvalue expressions become more cumbersome. Using the parameter values suggested in~\cite{Vorobev:2019}, we plot $\lambda_{1}$ and $\lambda_{2}$ as $R_c$ is scaled via $R_c=\alpha X_c$, $\alpha\in\{0,0.05,0.1,0.15,0.2\}$, for the matrix ${\bf M}\mathcal{Y}+({\bf M}\mathcal{Y})^{\dagger}$ (we plot admittance eigenvalues rather than impedance eigenvalues solely for graphical clarify). Fig. \ref{fig: Eig_Plot} shows the eigenvalues. Clearly, as coupling resistance decreases in value, the inverter behaves more passively. The DEF method will thus perform more successfully when virtual + coupling resistance is minimized.
\begin{figure}
\centering
\includegraphics[width=\columnwidth]{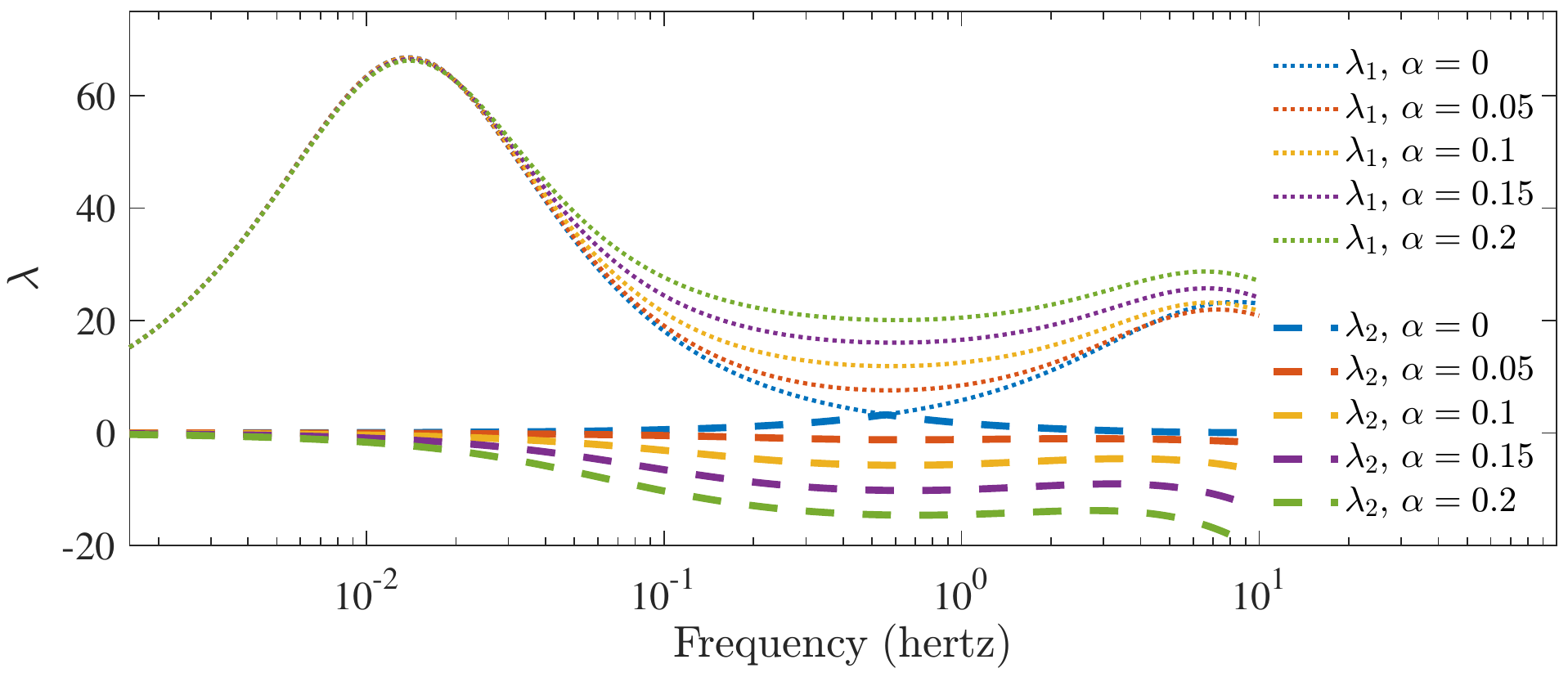}
\caption{\label{fig: Eig_Plot} {Plot of the eigenvalues of ${\bf M}\mathcal{Y}+({\bf M}\mathcal{Y})^{\dagger}$, where $\mathcal{Y}$ is associated with the dynamics of a droop-controlled inverter. For $\alpha=0$, the system is fully passive since both eigenvalues are positive.}}
\end{figure}}

\subsubsection{{Third Order Generator with First Order AVR}}
{We consider the dynamics of a third order synchronous machine with a first order ARV. The associated model~\cite{Milano:2013} is stated by
\begin{subequations}\label{eq: GenModel}
\begin{align}
\dot{\delta} & =\Delta\omega\\
M\Delta\dot{\omega} & =P_{m}-P_{e}-D\Delta\omega\\
T_{d0}'\dot{e}_{q}' & =E_{f}-(X_{d}-X_{d}')i_{d}-e_{q}'\\
T_{a}\dot{E}_{f} & =K_{a}({\rm V}_{r}-{\rm V})-E_{f},
\end{align}
\end{subequations}
where ${\rm V}_r$ is the voltage reference set point, ${\rm V}e^{j\theta}$ is the terminal voltage phasor, $i_d=(e_q'-e_q)/X_d'$, $i_q=e_d/X_q$, $e_d={\rm V}\sin(\delta-\theta)$ and $e_q={\rm V}\cos(\delta-\theta)$. Finally, $P_{e}=e_{d}i_{d}+i_{q}e_{q}$. We define output power variables $P$ and $Q$ as $P=-P_{e}$ and $Q=-(i_{d}e_{q}-i_{q}e_{d})$\footnote{{{In keeping with the conventions of Section \ref{Network Model}}, positive currents flow from the network and into the machine.}}.\\
\indent In order to make analytical claims about the passivity of (\ref{eq: GenModel}), we leverage the alternative DEF interpretation proposed in Appendix \ref{AppD}. To make the analysis tractable, we choose to linearize the model about the generator's unloaded equilibrium point (i.e. $\theta\!=\!\delta$, ${\rm V}\!=\!e_q$, $e_d\!=\!0$, etc.). Although this configuration may be uncommon, the resulting linearized dynamics are sufficiently simplified for approximating generator passivity. Additionally, the resulting admittance matrix $\mathcal H$ is diagonal, so the eigenvalues of ${\bf K}^{\dagger}\mathcal{H}+({\bf K}^{\dagger}\mathcal{H})^{\dagger}$ are trivial. We compute $\mathcal H$ from Appendix \ref{AppD}:
\begin{align}\label{eq: H_gen}
\mathcal{H} & \!=\!\!\left[\!\!\!\!\begin{array}{cc}
\frac{\Omega(jM\Omega+D)}{DX_{q}\Omega+j\left(MX_{q}\Omega^{2}-1\right)} \!\!\!&\!\!\! 0\\
0 \!\!\!&\!\!\! \frac{K_{a}+1-\Omega^{2}T_{a}T_{d0}'+j\Omega\left(T_{a}+T_{d0}'\right)}{(\Omega T_{a}-j)(X_{d}j-\Omega T_{d0}'X_{d}')}
\end{array}\!\!\!\!\right]\!.
\end{align}
Taking the eigenvalues of ${\bf K}^{\dagger}\mathcal{H}+({\bf K}^{\dagger}\mathcal{H})^{\dagger}$ is equivalent to taking the imaginary parts of the diagonals of matrix $\mathcal H$. The relationship between $\tilde P$ and $\tilde \theta$ (i.e. $\lambda_1$) is thus passive if
\begin{align}
\Omega D\ge 0.
\end{align}
This will be true if damping $D$ is positive. Interestingly, when the generator is \textit{unloaded}, the damping provided by the field winding is of second order and thus has no mathematical impact on passivity between $\tilde P$ and $\tilde \theta$. Via (\ref{eq: H_gen}), the relationship between $\tilde Q$ and $\tilde {\rm V}$ (i.e. $\lambda_2$) is thus passive if
\begin{align}
\left(T_{d0}'\Omega^{2}T_{a}^{2}\!+\!T_{d0}'\right)\left(X_{d}\!-\!X_{d}'\right)\!-\!K_{a}\!\left(T_{d0}'X_{d}'\!+\!X_{d}T_{a}\right)\!\ge\! 0.
\end{align}
This result has many interesting interpretations. For $K_a$ small or $T_a$ sufficiently large, the expression will always be passive. Strong and fast AVR response thus causes the relationship between $\tilde Q$ and $\tilde {\rm V}$ to lose passivity. For passivity to be guaranteed $\forall \Omega$, the AVR gain should have the upper bound
\begin{align}\label{eq: passivity_bound}
K_{a} & \le\frac{T_{d0}'(X_{d}-X_{d}')}{T_{d0}'X_{d}'+X_{d}T_{a}}.
\end{align}
Of course, the claim that the generator will be passive if $D\ge0$ and (\ref{eq: passivity_bound}) are both satisfied is only true when the generator is unloaded. Loading effects are likely to be small, though, so these results are useful for guiding intuition related to when generators will lose passivity.}


\section{Test Results}\label{Test Results}
{In this section, we present test results which illustratively validate the framework presented in Sections \ref{Network Model} and \ref{Energy Function Analysis} and the practical applications presented in Section \ref{Practical Applications}. We perform these tests in two systems which are altered to engender poor DEF performance. In the 39-bus New England system, the average R/X transmission line ratio was increased from $6\%$ to $15\%$, and in the 179-bus WECC system, all constant power loads were converted into constant impedance loads. All simulation code is posted online\footnote{https://github.com/SamChevalier/Passivity-Enforcement-FOs} for open source access.}

\subsection{{New England 39-bus Test System}}
\begin{figure}
\centering
\includegraphics[scale=0.48]{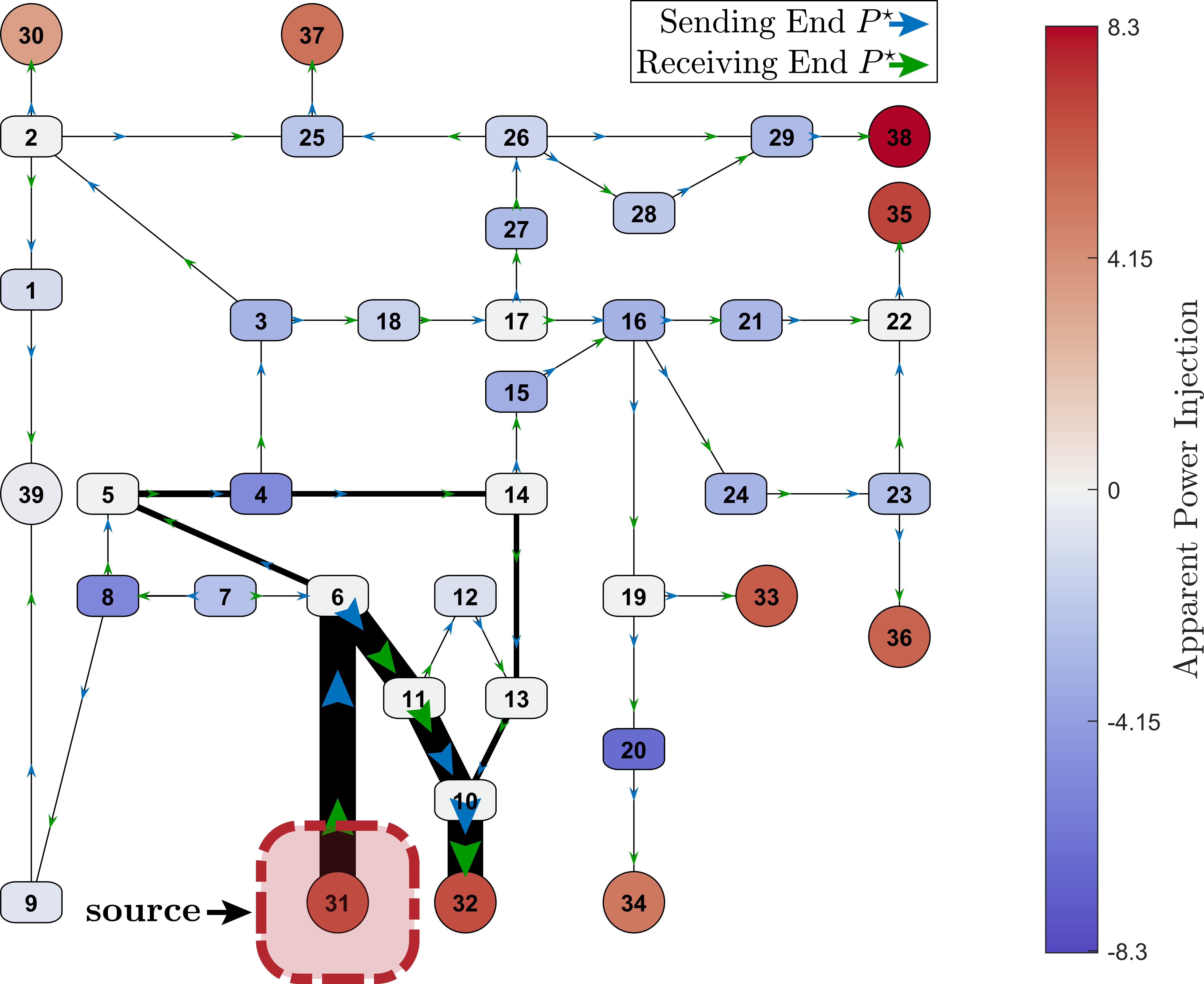}
\caption{\label{fig: Full_Sys_nR} Dissipating power flow $P^{\star}$ in the IEEE 39-bus power system {for an applied FO of 2 Hz. All line resistance has been removed.} Circles represent generators while rounded rectangles represent loads. Line thickness and arrow size represent dissipating power flow magnitude.}
\end{figure}

{Fig. \ref{fig: Full_Sys_nR} shows a diagram of the IEEE 39-bus New England system. In this system, generators are modeled as third order synchronous machines~\cite{Sauer:2006} with first order AVRs\footnote{The gains and time constants of these AVRs were chosen to approximate the full regulator + exciter system model from~\cite{Sauer:2006}.}. Initially, loads were modeled as constant power (PQ) and system lines were modeled with no-loss (purely reactive). Following the steps in Algorithm \ref{alg: DEF_Check}, we computed the eigenvalues associated with bus 31's DWE ${\mathcal Y}_N$ for a frequency of $\Omega_d=2\pi\times2$:
\begin{align}\label{eq: pos_lossless}
    \lambda\left\{{\bf M}\mathcal{Y}_{N}+({\bf M}\mathcal{Y}_{N})^{\dagger}\right\}=0.007,\,2.11.
\end{align}
Due to the two positive eigenvalues, any FO source originating at bus 31 with a frequency of 2 Hz will appear non- passive: the DEF method should perform accurately in this system since condition (\ref{eq: damping_condition}) is satisfied. We then simulated the response of the system to a 2Hz perturbation originating at bus 31. System excitation and simulation were performed in the frequency domain. From the simulated data, dissipating power $P^{\star}$ was computed via (\ref{eq: DEF_passivity}) on all lines. The resulting flows are plotted in Fig. \ref{fig: Full_Sys_nR}. Clearly, the source is readily identifiable.}

\begin{figure}
\centering
\includegraphics[scale=0.48]{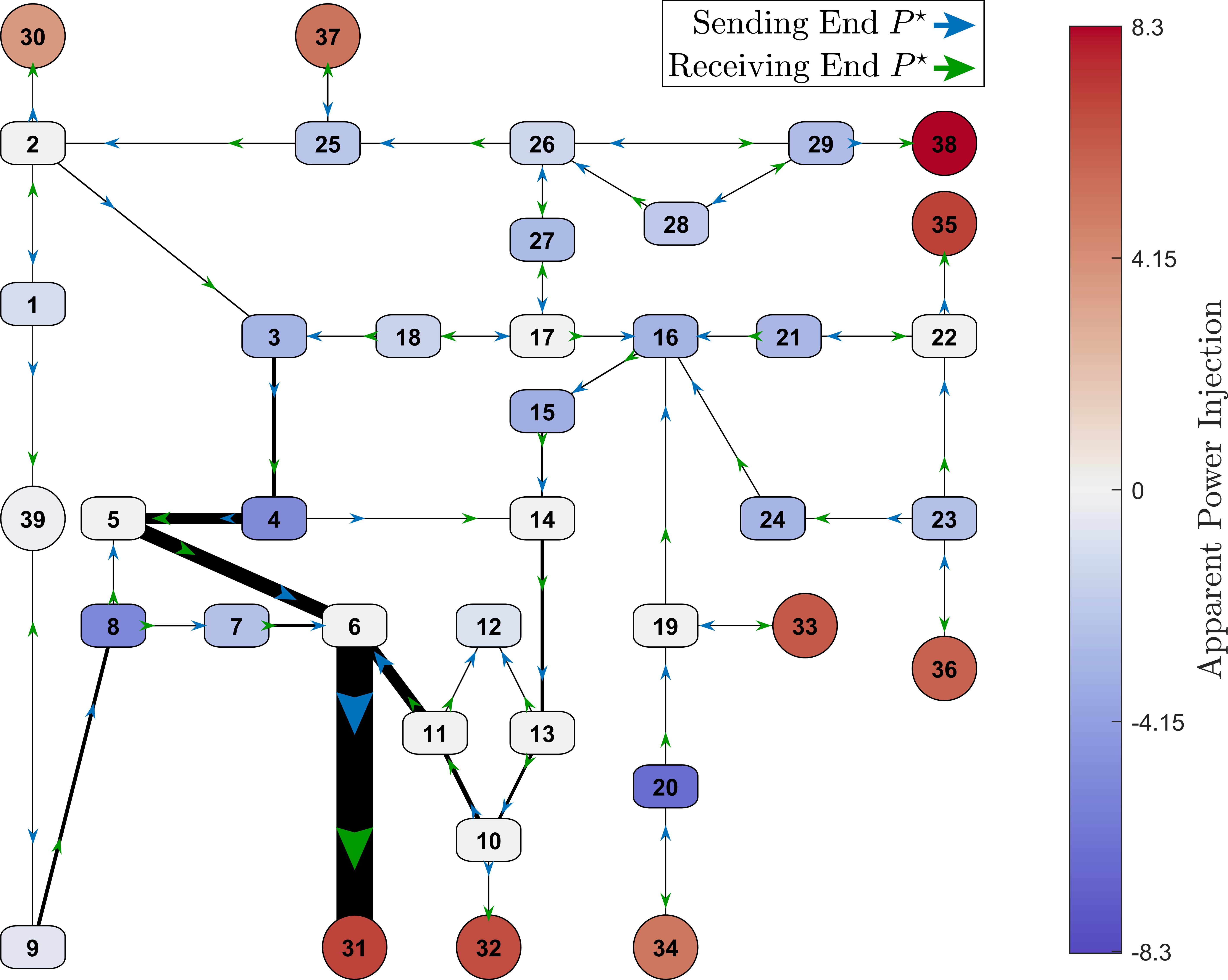}
\caption{\label{fig: Full_Sys} Dissipating power flow $P^{\star}$ in the lossy IEEE 39-bus power system. {The source of the FO cannot be located using the energy-based DEF method in this situation.}}
\end{figure}

{Next, the model was altered such that the average\footnote{{One outlier line, which has $R\approx X$, was excluded from this average.}} R/X ratio was increased from $6\%$ to $15\%$ for all lines. Additionally, all loads were converted to the model described by (\ref{eq: PQ}) with $\alpha$ and $\beta$ parameters chosen randomly from ${\mathcal U}(0,2)$. We again considered the eigenvalues associated with bus 31's DWE ${\mathcal Y}_N$ for a frequency of $\Omega_d=2\pi\cdot2$:
\begin{align}
    \lambda\left\{{\bf M}\mathcal{Y}_{N}+({\bf M}\mathcal{Y}_{N})^{\dagger}\right\}=-2.52,\,+4.77
\end{align}
The oppositely signed eigenvalues show that the network can aggregately generate or consume dissipating power $P^{\star}$, indicating unreliable DEF results.} After computing the system's response to a 2Hz sinusoidal FO applied at bus 31's generator, the dissipating power $P^{\star}$ flowing in the network was plotted in Fig. \ref{fig: Full_Sys}. Based on arrow directionality, all generators are shown to be $P^{\star}$ \textit{sinks}: there is no apparent generator source in the system. {Loads present only a small contribution of dissipating power. Although it may appear as though bus 6, for example, is a source, the dissipating power flows are exactly conserved at this bus: all dissipating power flowing in from lines 7-6, 5-6, and 11-6 exactly flows out on line 6-31.}

The reason there are no apparent sources is due to the highly active nature of the lines. For example, in the flow from bus 11 to bus 6, the sending end $P^{\star}$ is larger than the receiving end $P^{\star}$ (based on arrow sizes), and given the flow direction, the line is clearly a source of dissipating power. Similarly, on line 21-22, both arrows point away from the line, indicating positive dissipating power flows out of both ends of the line. {Becasue condition (\ref{eq: damping_condition}) is not met in this network, the FO source is able to act as a dissipating power sink. In other words, since the network is producing more dissipating power than it can consume, the source provides additional ``sink" slack.} Therefore, the source cannot be readily identified by a system operator.

{As a second test, we reconsidered the no-loss system whose eigenvalues are characterized by (\ref{eq: pos_lossless}). At load buses 4, 8 and 16, we added identical droop-controlled inverter circuits whose parameter values are specified in~\cite{Vorobev:2019}. However, we intentionally neglected resistive losses (i.e. $R_c=0$). Next, we scaled the nominal droop gain values of $k_p=1.3\cdot10^{-3}$ and $k_q=7.5\cdot10^{-3}$ by scalar $\alpha\in[0 \,\; 10^4]$. For each system configuration, the eigenvalues of the DWE seen at the source bus were computed per Algorithm \ref{alg: DEF_Check}. The results, shown in Table \ref{table: droop_evals}, confirm the prediction of (\ref{eq: droop_evals}). Across the full range of droop gain values, the eigenvalues of the source bus are both positive. Therefore, these components do not interfere with DEF performance. When losses are added back in, interference may occur.}
\begin{table}
\renewcommand{\arraystretch}{1}
\caption{{Eigenvalues of System with Droop-Controlled Inverters}}\label{table: droop_evals}
\centering\label{tab:vars}
\begin{tabular}{|c|c|c|c|c|c|c|}
\hline
&  $\alpha\!=\!0$ & $\alpha\!=\!10^0$ & $\alpha\!=\!10^1$ & $\alpha\!=\!10^2$ & $\alpha\!=\!10^3$ & $\alpha\!=\!10^4$\\
\hline
\hline
$\lambda_1$       & 0.007 & 0.007 & 0.008 & 0.01 & 0.05 & 0.4\\
$\lambda_2$       & 2.11  & 2.11  & 2.12  & 2.17 & 2.80 & 28.04\\
\hline
\end{tabular}
\end{table}

{As a final test on the 39-bus system, we {again} reconsidered the no-loss system whose eigenvalues are characterized by (\ref{eq: pos_lossless}). Instead of adding loss, we tripled all AVR gain values and decreased all AVR time constants by one third. Accordingly,
\begin{align}
    \lambda\left\{{\bf M}\mathcal{Y}_{N}+({\bf M}\mathcal{Y}_{N})^{\dagger}\right\}=-0.1,\,+1.53.
\end{align}
As evidenced by the slightly negative eigenvalue, these changes caused generator passivity to be lost in at least some cases. This effect was predicted by (\ref{eq: passivity_bound}).}

\subsection{{WECC 179-bus Test System}}\label{WECC 179-bus Test System}
\begin{figure}
\centering
\includegraphics[width=\columnwidth]{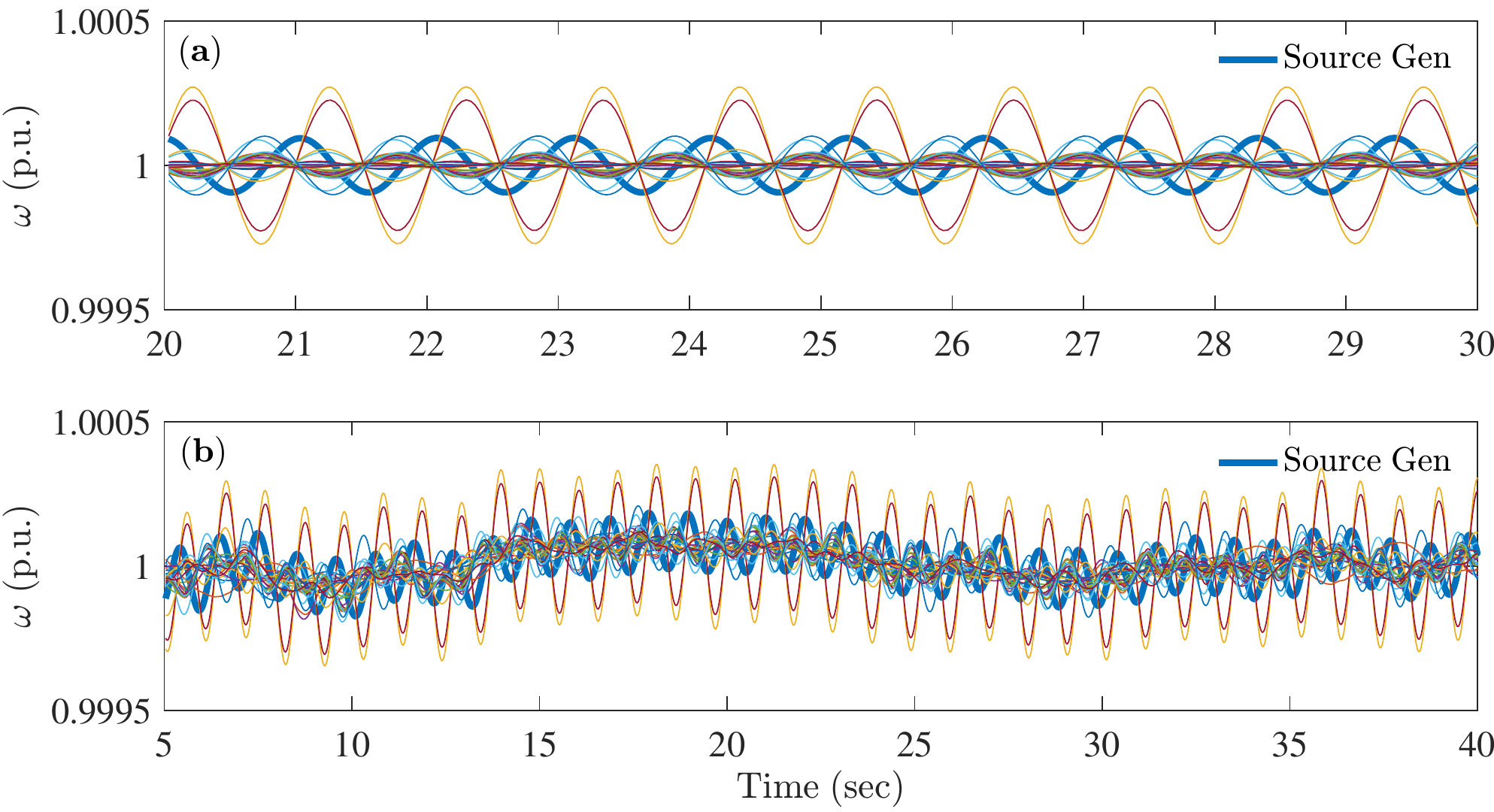}
\caption{\label{fig: Frequency_Oscillations}{Shown are the generator frequency oscillations for all 29 system generators. Panel $({\bf a})$ shows 10s of time series data for when the FO is the only source of system excitation, and panel $({\bf b})$ shows 35s of time series data for when the FO and random load perturbations are both exciting the system.}}
\end{figure}

{In the second test, we employed the 179-bus WECC system which was prepared by the IEEE Task Force on FOs~\cite{Maslennikov:2016TF}. Specifically, we employed test case number F1, where a 0.86 Hz FO is applied to the reference signal in a generator's AVR. We altered the system by converting all loads to constant impedance with some frequency dependence. We also applied Ornstein-Uhlenbeck noise on the load power terms $P_0$ and $Q_0$ from (\ref{eq: P(t)})-(\ref{eq: Q(t)}). {Due to the load model modification, the natural frequencies of the system changed slightly. Thus, i}n order to engender better resonance, we increased the frequency of the FO by 0.1Hz, and we increased the damping at the source generator to ensure it acted as a FO sink in the DEF analysis. We simulated the system for 120 seconds with Power System Analysis Toolbox (PSAT)~\cite{Milano:2013}. To show the strength of the resonance condition, we initially simulated the system with no load noise and plotted the time domain frequency response $\omega$ of the system generators. Panel $({\bf a})$ of Fig. \ref{fig: Frequency_Oscillations} shows that there are multiple generators with larger frequency oscillations than the source. Panel $({\bf b})$ shows the frequency dynamics of the system generators once load stochasticisticy is added.}

\begin{figure}
\centering
\includegraphics[width=\columnwidth]{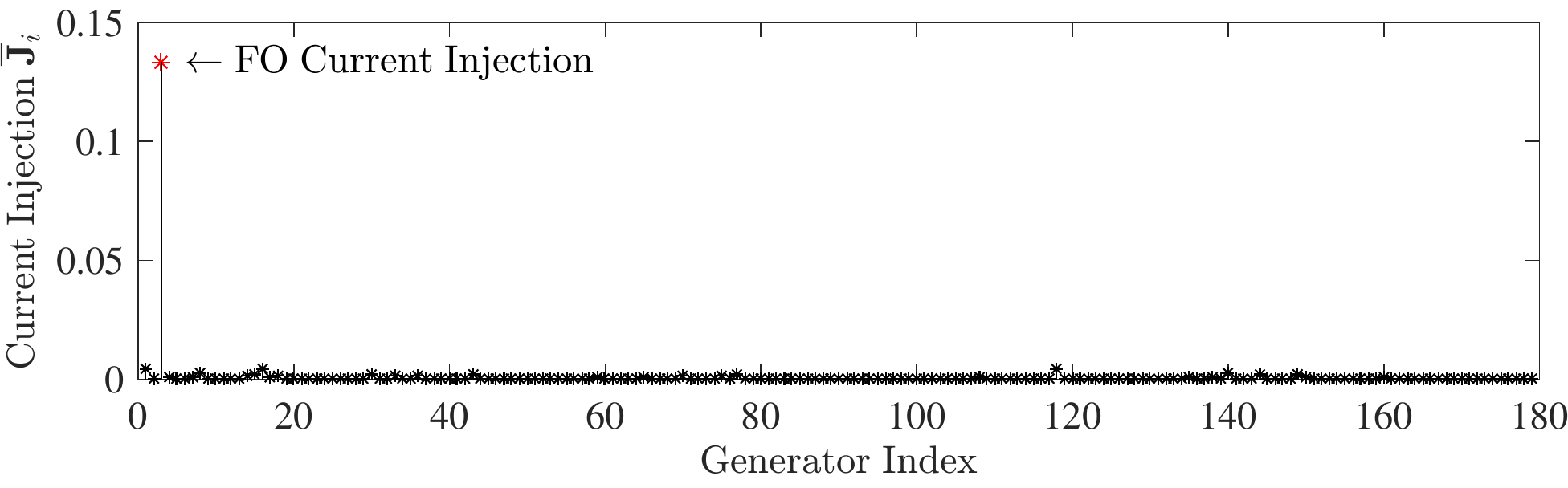}
\caption{\label{fig: Current_Injs}{Plotted are the current injection magnitudes of (\ref{eq: J_mag}). The perturbative model (\ref{eq: J=YV_full}) predicts nodal current balances with a high degree of accuracy. Non-source current injections are due to extraneous load perturbations.}}
\end{figure}

{Before investigating the dissipating energy flows in this large system, we first sought to further experimentally validate the propagation framework proposed in Section \ref{Network Model}. To do so, we analytically constructed the full system model of (\ref{eq: Yb_Model}). As predicted by this model, the complex current injection vector ${\bf J}\!\in\!{\mathbb C}^{2n\times1}$ from (\ref{eq: J=YV_full}) should be sparse: non-zero elements should only come from unmodeled, extraneous perturbations. We thus simulated the WECC system with noisy loads and a strong FO source at bus 4. We then collected all voltage PMU data, generated the FFT evaluated at $\Omega_d=2\pi\times0.96$, constructed ${\bf V}_b$ from (\ref{eq: J=YV_full}) and the methods outlined in Appendix \ref{AppB}, and predicted the current injection vector $\bf J$ via ${\bf J} =-\mathcal{Y}_{B}{\bf V}_{b}$. Since there are two complex current injection components associated with each bus, we defined injection magnitude vector $\overline{\bf J}\in {\mathbb R}^{n\times 1}$. Its $i^{\rm th}$ element is given as the sum of the magnitudes of the current injections at the $i^{\rm th}$ bus:
\begin{align}\label{eq: J_mag}
    \overline{{\bf J}}_{i}=|{\bf J}_{2i-1}|+|{\bf J}_{2i}|.
\end{align}}
{We plot the injection magnitudes in Fig. \ref{fig: Current_Injs}. As expected, the primary current injection is located at the source bus, but there is a plethora of small, non-zero injections as well. These come from the random load perturbations which are applied at each time step. Fig. \ref{fig: Current_Injs} thus serves to validate (\ref{eq: Yb_Model}) as a useful model for analyzing FO propagation.}

\begin{figure}
\centering
\includegraphics[width=\columnwidth]{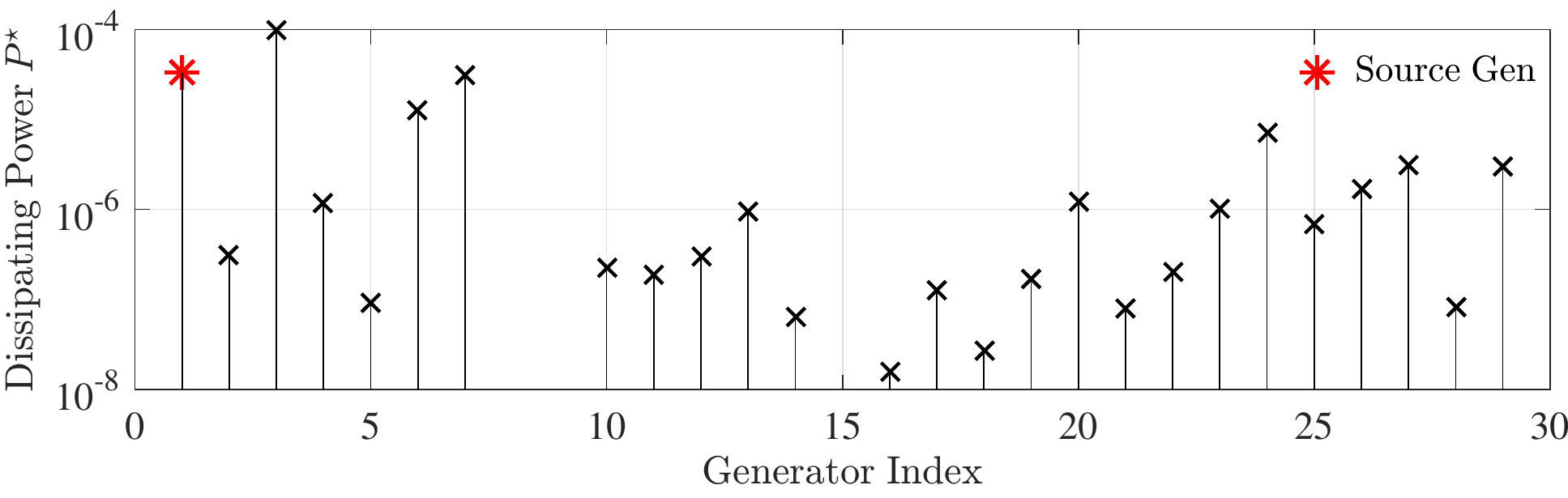}
\caption{\label{fig: DEF_Gens}{Plotted are the dissipating power injections across all 29 generators. All injections are positive, indicating the positive dissipation of $P^{\star}$. Generators are listed in ascending order of corresponding bus number. Some magnitudes are too small to appear on the plot.}}
\end{figure}

{As stated, the FO source is located in the generator at bus 4. By converting all loads to constant impedance (with some frequency dependence), we were able to engender a system whose DWE ${\mathcal Y}_N$ had oppositely signed eigenvalues. Via Algorithm \ref{alg: DEF_Check},
\begin{align}\label{eq: eigs_WECC}
\lambda\left\{{\bf M}\mathcal{Y}_{N}+({\bf M}\mathcal{Y}_{N})^{\dagger}\right\}=-11.9,\,+39.6.
\end{align}
Again, these are the eigenvalues seen by the source bus when it looked into the system. Because one of the eigenvalues is negative, the source generator is capable of acting as a dissipating power sink. When the AVR oscillation was initially applied, simulation results showed that the generator acted like a source. When we simply increased the damping of the source generator though, we began to excite alternative eigenvalues of ${\mathcal Y}_N$, and the source generator became a sink. For a sufficiently high level of source damping, the dissipating power injections of Fig. \ref{fig: DEF_Gens} were observed at the generator buses. In this scenario, the DEF method fails due to the high level of resistance in the system. This unreliable DEF performance is predicted by the oppositely signed eigenvalues of (\ref{eq: eigs_WECC}).}

{It is instructive to note that the eigenvalues of (\ref{eq: eigs_WECC}) do not change as the damping at the source bus is altered. These eigenvalues are a product of the network, not the source. This is further confirmation that Algorithm \ref{alg: DEF_Check} is agnostic to the type or cause of the FO; rather, Algorithm \ref{alg: DEF_Check} predicts the network response to any perturbation originating from the selected source bus. Accordingly, frequency and topological location are the only important characteristics of the FO considered in the algorithm.}

\section{{Conclusion and Future Work}}\label{Conclusion}
{In this paper, we have proposed a novel framework for understanding how FOs linearly propagate in an electrical power network. Using Tellegan's theorem and passivity concepts, we leveraged this perturbative propagation framework to further motivate and analyze the successful DEF method. We proved that the DEF's shortcomings in a classical power system cannot be avoided by simply selecting a new passivity transformation (quadratic energy function), and in Theorem \ref{theorem4}, we showed that the effective source admittance associated with a FO cannot be passive in a classical power system. We further developed necessary (and by extension, sufficient) conditions for the failure of the DEF method. When power system operators have access to an analytical system model, they may use our proposed framework to analytically predict apriori if the DEF method might fail ($\lambda_1\ge 0$, $\lambda_2\le 0$), will fail ($\lambda_1,\lambda_2\le 0$), or will succeed ($\lambda_1,\lambda_2\ge 0$). This may be especially useful in small, microgrid networks where system architecture is known more fully and line loss ratios are much higher.}

{One obvious drawback to the proposed approach for the apriori testing of DEF performance is the need for an analytical system model. Load models, especially, may be highly uncertain from the perspective of a system operator. And as observed in subsection \ref{Inf_Analysis}, inference of the underlying (load or generator) admittances cannot be performed from a single PMU observation due to the underdetermined nature of the proplem. In future work, we hope to propose model-free and data-driven inference techniques which overcome this issue by leveraging surrounding spectral data to regularize the inference problem. Although predicting DEF performance is an important contribution in itself, future research will also focus on leveraging this framework to devise methods which enhance the source location performance of energy-based methods.}

{The main theorems in this paper assumed a classical lossy power system, but the underlying framework presented in Sections \ref{Network Model} and \ref{Energy Function Analysis} is actually quite general and can be used to interpret a variety of linear propagation phenomena. As a further extension of this work, we hope to use our framework to investigate the passivity properties of other renewable energy sources, such as doubly-fed induction generators (DFIGs) and power electronic interfaces of HVDC lines. We also plan to use these approaches to analyze the performance of the DEF method in the case of interarea oscillations{, especially those caused by the interaction of many positively damped AVRs. The performance of energy-based oscillation detection methods in this situation is still an open research question.}}

\appendices
{\section{}\label{AppA}
We consider the complications associated with a two port element whose current flow is not a linear function of terminal voltage differentials. To do so, we return to a \textit{standard} steady state (rather then perturbative) power system model, and we consider a transmission line with complex admittance ${\tilde y}_{ij}\in{\mathbb C}^1$. If ${\tilde V}_i$, ${\tilde V}_j\in{\mathbb C}^1$ are the phasor voltages associated with buses $i$ and $j$, then the standard two port model associated with a tap changing transformer may be stated as
\begin{align}
\left[\begin{array}{c}
\tilde{I}_{ij}\\
\tilde{I}_{ji}
\end{array}\right]=\left[\begin{array}{cc}
\frac{\tilde{y}_{ij}}{c^{2}} & -\frac{\tilde{y}_{ij}}{c}\\
-\frac{\tilde{y}_{ij}}{c} & \tilde{y}_{ij}
\end{array}\right]\left[\begin{array}{c}
\tilde{V}_{i}\\
\tilde{V}_{j}
\end{array}\right],
\end{align}
where $c$ is the tap ratio and ${\tilde I}_{ij}$, ${\tilde I}_{ji}\in{\mathbb C}^1$ are the current flows. Considering the top equation, the flow $\tilde{I}_{ij}$ may be written as
\begin{subequations}
\begin{align}
\tilde{I}_{ij} & =\frac{\tilde{y}_{ij}}{c^{2}}\tilde{V}_{i}-\frac{\tilde{y}_{ij}}{c}\tilde{V}_{j}\\
 & =\frac{\tilde{y}_{ij}}{c}\left(\tilde{V}_{i}-\tilde{V}_{j}\right)+\tilde{V}_{i}\tilde{y}_{ij}\left(\frac{1-c}{c^{2}}\right).\label{eq: flow_decomp}
\end{align}
\end{subequations}
From (\ref{eq: flow_decomp}), we may thus interpret ${\tilde{y}_{ij}}/c$ as the series impedance between the lines, and we may interpret $\tilde{V}_{i}\tilde{y}_{ij}({1-c})/{c^{2}}$ as a \textit{shunt} current injection at bus $i$ caused by the tap changer. A similar shunt injection will exist at bus $j$. In this way, any two port element whose current flow isn't directly proportional to $\tilde{V}_{i}-\tilde{V}_{j}$ (phase shifting transformer, HVDC line, etc.) can be expressed as the sum of a linear flow term and a shunt injection term. Once these relations are appropriately converted into the perturbative system model, matrices (\ref{eq: YL_mat}) and (\ref{eq: YS_mat}) can be updated accordingly.
\section{}\label{AppB}
An inherent problem exists when attempting to compute (\ref{eq: DEF_passivity}) in a realistic power system. Because the system has natural integration effects, the phase angles of the system constantly drift; this can be seen in the (simulated or real) time series data of any system with stochastic loads. Because generator admittance can be a function of steady state rotor angle $\delta$, such as in (\ref{eq: Yg}), the underlying admittance will be constantly changing as system phase angle drifts. Due to the infusion of such nonlinear effects, FFT analysis of rectangular (or Cartesian) coordinate variables will yield outputs which are incompatible with the perturbative model proposed in Section \ref{Network Model}: the underlying admittances will be constantly changing as system phase angle drifts. Polar coordinate formulations, however, do not have this same drawback. As a workaround, we may introduce the concept of \textit{effective} rectangular perturbation. To do so, we first introduce matrices ${\rm T}_1$ and ${\rm T}_{1,i}$ which linearize standard phasors $V_{r}+jV_{i}={\rm V}e^{j\theta}$ and $I_{r}+jI_{i}={\rm I}e^{j\phi}$, respectively~\cite[eq. (10)]{Chevalier:2018}:
\begin{align}
\left[\begin{array}{c}\label{eq: T1_mat}
\Delta V_{r}\\
\Delta V_{i}
\end{array}\right] & =\underbrace{\left[\begin{array}{cc}
\cos(\theta) & -{\rm V}\sin(\theta)\\
\sin(\theta) & {\rm V}\cos(\theta)
\end{array}\right]}_{{\rm T}_{1}}\left[\begin{array}{c}
\Delta{\rm V}\\
\Delta\theta
\end{array}\right]\\
\left[\begin{array}{c}\label{eq: T1i_mat}
\Delta I_{r}\\
\Delta I_{i}
\end{array}\right] & =\underbrace{\left[\begin{array}{cc}
\cos(\phi) & -{\rm I}\sin(\phi)\\
\sin(\phi) & {\rm I}\cos(\phi)
\end{array}\right]}_{{\rm T}_{1,i}}\left[\begin{array}{c}
\Delta{\rm I}\\
\Delta\phi
\end{array}\right].
\end{align}
We now assume we have some admittance ${\mathcal Y}_g$ which relates polar voltage and current perturbations via $\tilde{{\bf I}}_p=\mathcal{Y}_{g}\tilde{{\bf V}}_p$. This relationship will not be affected by system phase angle drift. We thus transform the expression via
\begin{subequations}
\begin{align}
{\rm T}_{1,i}\tilde{{\bf I}}_{p}= & {\rm T}_{1,i}\mathcal{Y}_{g}{\rm T}_{1}^{-1}{\rm T}_{1}\tilde{{\bf V}}_{p}\\
\tilde{{\bf I}}_{e}= & \left({\rm T}_{1,i}\mathcal{Y}_{g}{\rm T}_{1}^{-1}\right)\tilde{{\bf V}}_{e},\label{eq: Ie_Ve}
\end{align}
\end{subequations}
where $\theta$ and $\phi$ from (\ref{eq: T1_mat})-(\ref{eq: T1i_mat}) are chosen as their respective instantaneous values at $t\!=\!0$ in the PMU data (any reference angle may be used so long as it is consistent across all transformations). Vectors $\tilde{{\bf I}}_{e}$ and $\tilde{{\bf V}}_{e}$ represent the effective rectangular perturbation vectors. Thus, (\ref{eq: Ie_Ve}) is a fully consistent equation and may be used to update the $P^{\star}$ flow equation in (\ref{eq: DEF_passivity}).}
{\section{}\label{AppC}}
{Rather than dealing with the admittance $\mathcal Y$ from ${\tilde {\bf I}} = {\mathcal Y}{\tilde {\bf V}}$, it may be convenient to deal with impedance ${\mathcal Z}={\mathcal Y}^{-1}$ instead. We define ${\bf N}_{\mathcal{Y}} ={\bf M}\mathcal{Y}+({\bf M}\mathcal{Y})^{\dagger}$ and ${\bf N}_{\mathcal{Z}} ={\bf M}\mathcal{Z}+({\bf M}\mathcal{Z})^{\dagger}$.
\begin{theorem}
Iff ${\bf N}_{\mathcal{Y}}$ is (non)passive, ${\bf N}_{\mathcal{Z}}$ is (non)passive.
\begin{proof}
Starting with $\tilde{{\bf I}}=\mathcal{Y}\tilde{{\bf V}}$, we multiply through by $\tilde{{\bf V}}^{\dagger}{\bf M}$:
\begin{align}
    {\rm Re}\{\tilde{{\bf V}}^{\dagger}{\bf M}\tilde{{\bf I}}\}&={\rm Re}\{\tilde{{\bf V}}^{\dagger}{\bf M}\mathcal{Y}\tilde{{\bf V}}\}.
\end{align}
Similarly, starting with $\mathcal{Z}\tilde{{\bf I}} =\tilde{{\bf V}}$, we multiply through by $\tilde{{\bf I}}^{\dagger}{\bf M}$:
\begin{align}
   {\rm Re}\{\tilde{{\bf I}}^{\dagger}{\bf M}\mathcal{Z}\tilde{{\bf I}}\} & ={\rm Re}\{\tilde{{\bf I}}^{\dagger}{\bf M}\tilde{{\bf V}}\}.
\end{align}
Since ${\rm Re}\{\tilde{{\bf V}}^{\dagger}{\bf M}\tilde{{\bf I}}\}=\alpha\in\mathbb R$, then
\begin{subequations}
\begin{align}
    (\alpha)^{\dagger} & ={\rm Re}\{\tilde{{\bf V}}^{\dagger}{\bf M}\tilde{{\bf I}}\}^{\dagger}\\
 & ={\rm Re}\{\tilde{{\bf I}^{\dagger}}{\bf M}\tilde{{\bf V}}\}
\end{align}
\end{subequations}
since ${\bf M}={\bf M}^{\dagger}$. Because ${\rm Re}\{\tilde{{\bf I}}^{\dagger}{\bf M}\mathcal{Z}\tilde{{\bf I}}\} ={\rm Re}\{\tilde{{\bf V}}^{\dagger}{\bf M}\mathcal{Y}\tilde{{\bf V}}\}$,
\begin{align}
\tilde{{\bf I}}^{\dagger}({\bf M}\mathcal{Z}+({\bf M}\mathcal{Z})^{\dagger})\tilde{{\bf I}} & =\tilde{{\bf V}}^{\dagger}({\bf M}\mathcal{Y}+({\bf M}\mathcal{Y})^{\dagger})\tilde{{\bf V}}.
\end{align}
Thus, ${\bf N}_{\mathcal{Y}}$ and ${\bf N}_{\mathcal{Z}}$ share the same passivity classification.
\end{proof}
\end{theorem}}
{\section{}\label{AppD}}
{The original DEF integral, as given in~\cite{Chen:2013}, may be stated as
\begin{subequations}\label{eq: W_DE}
\begin{align}
W_{{\rm DE}} & =\int{\rm Im}\left\{ I^{*}{\rm d}V\right\} \\
 & =\int{\rm Im}\left\{ \left(\tfrac{P+jQ}{{\rm V}e^{j\theta}}\right){\rm d}{\rm V}e^{j\theta}\right\}
\end{align}
\end{subequations}
where ${\rm d}{\rm V}e^{j\theta}=(\dot{{\rm V}}+j{\dot \theta}{\rm V})e^{j\theta}{\rm d}t$. We thus restate $W_{{\rm DE}}$ as
\begin{subequations}
\begin{align}
W_{{\rm DE}} & =\int{\rm Im}\{ \tfrac{1}{{\rm V}}({P+jQ})(\dot{{\rm V}}+j{\dot \theta}{\rm V}){\rm d}t\} \\
 & =\int(P{\dot \theta}+Q{\dot{{\rm V}}}/{\rm V}){\rm d}t.
\end{align}
\end{subequations}
Other publications commonly state the term $(Q{\dot{{\rm V}}}/{\rm V}){\rm d}t$ as $Q{\rm d}\ln({\rm V})$. Assuming small perturbations of (\ref{eq: W_DE}), we may write ${\rm d}\ln({\rm V}) \approx {\rm d}{\rm V}/{\rm V}_{0}=({\dot {\rm V}}/{\rm V}_0){\rm d}t$. We have
\begin{align}
W_{{\rm DE}} & \approx\int(P{\dot \theta}+Q{\dot{{\rm V}}}/{\rm V}_0){\rm d}t.
\end{align}
We now suppose there is some FRF $\mathcal H$ which relates output oscillations $\tilde P$ and $\tilde Q$ with input oscillations $\tilde {\rm V}$ and $\tilde \theta$ via
\begin{align}\label{eq: H_FRF}
\left[\begin{array}{c}
\tilde{P}\\
\tilde{Q}
\end{array}\right]=\mathcal{H}\left[\begin{array}{c}
\tilde{\theta}\\
\tilde{{\rm V}}
\end{array}\right].
\end{align}
We define a new passivity transformation matrix
\begin{align}
{\bf K}=\left[\!\begin{array}{cc}
j & 0\\
0 & \tfrac{j}{{\rm V}_{0}}
\end{array}\!\right].
\end{align}
We use $\bf K$ to write $\tilde{{\bf S}}=\mathcal{H}({\bf K}^{-1}{\bf K})\tilde{{\bf V}}_{p}$. The passivity of the underlying system may be tested via the following relation, where ${\tilde {\bf S}}$ and ${\tilde {\bf V}}_p$ are from (\ref{eq: H_FRF}) as previously defined in (\ref{eq: Ya}):
\begin{align}
{\rm Re}\{({\bf K}\tilde{{\bf V}}_{p})^{\dagger}\tilde{{\bf S}}\} & ={\rm Re}\{\tilde{{\bf V}}_{p}^{\dagger}({\bf K}^{\dagger}\mathcal{H})\tilde{{\bf V}}_{p}\}.
\end{align}
By the proof logic presented in~\cite{Chevalier:2019P}, the following conditions are equivalent after a sufficient number of perturbation cycles:
\begin{subequations}
\begin{align}
{\bf K}^{\dagger}\mathcal{H}+({\bf K}^{\dagger}\mathcal{H})^{\dagger}\succeq0 & \Leftrightarrow {\rm Re}\{\tilde{{\bf V}}_{p}^{\dagger}{\bf K}^{\dagger}\tilde{{\bf S}}\}\ge0\\
 & \Leftrightarrow W_{{\rm DE}}\ge0.\label{eq: WDE>0}
\end{align}
\end{subequations}
If instead of $\mathcal H$, the dynamics of the system were described by the admittance $\mathcal Y$, (\ref{eq: passivity_relations}) follows directly from (\ref{eq: WDE>0}):
\begin{align}\label{eq: passivity_relations}
{\bf K}\mathcal{H}+({\bf K}\mathcal{H})^{\dagger}\succeq0 &\,\Leftrightarrow\, {\bf M}\mathcal{Y}+({\bf M}\mathcal{Y})^{\dagger}\succeq0.
\end{align}}


\section*{{Acknowledgment}}
{The authors gratefully acknowledge the helpful correspondence and general guidance provided by Slava Maslennikov of ISO New England. ISONE's efforts, under Slava's direction, to pioneer the testing and implementation of the DEF method are to be sincerely applauded.}

\bibliographystyle{IEEEtran}
\bibliography{FO_Bib}

\begin{IEEEbiography}[{\includegraphics[width=1in,height=1.294in]{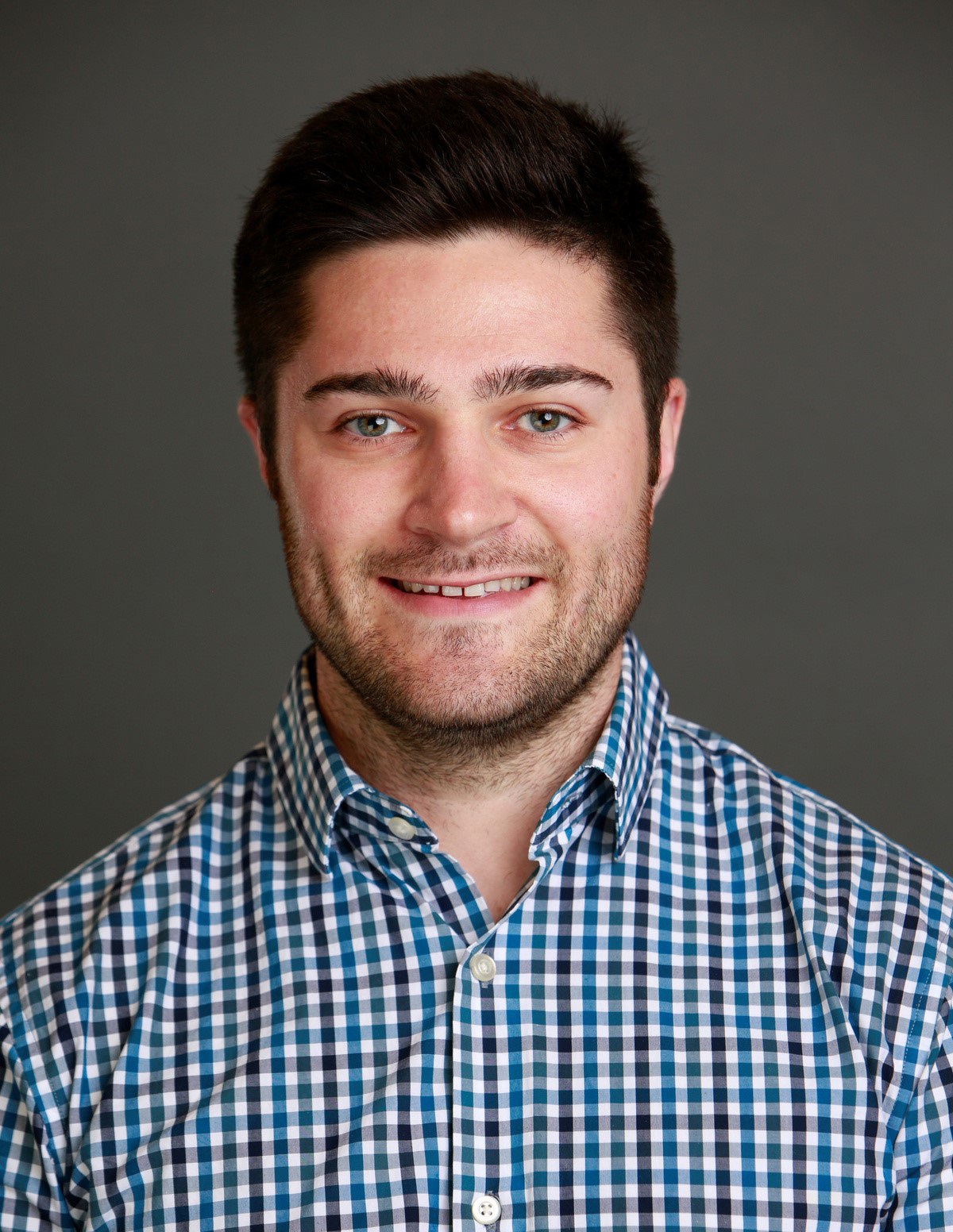}}]{{Samuel C.~Chevalier}} (S`13) received M.S. (2016) and B.S. (2015) degrees in Electrical Engineering from the University of Vermont, and he is currently pursuing the Ph.D. in Mechanical Engineering from the Massachusetts Institute of Technology (MIT). His research interests include electrical power system stability assessment and PMU applications.
\end{IEEEbiography}

\begin{IEEEbiography}[{\includegraphics[width=1in,height=1.294in]{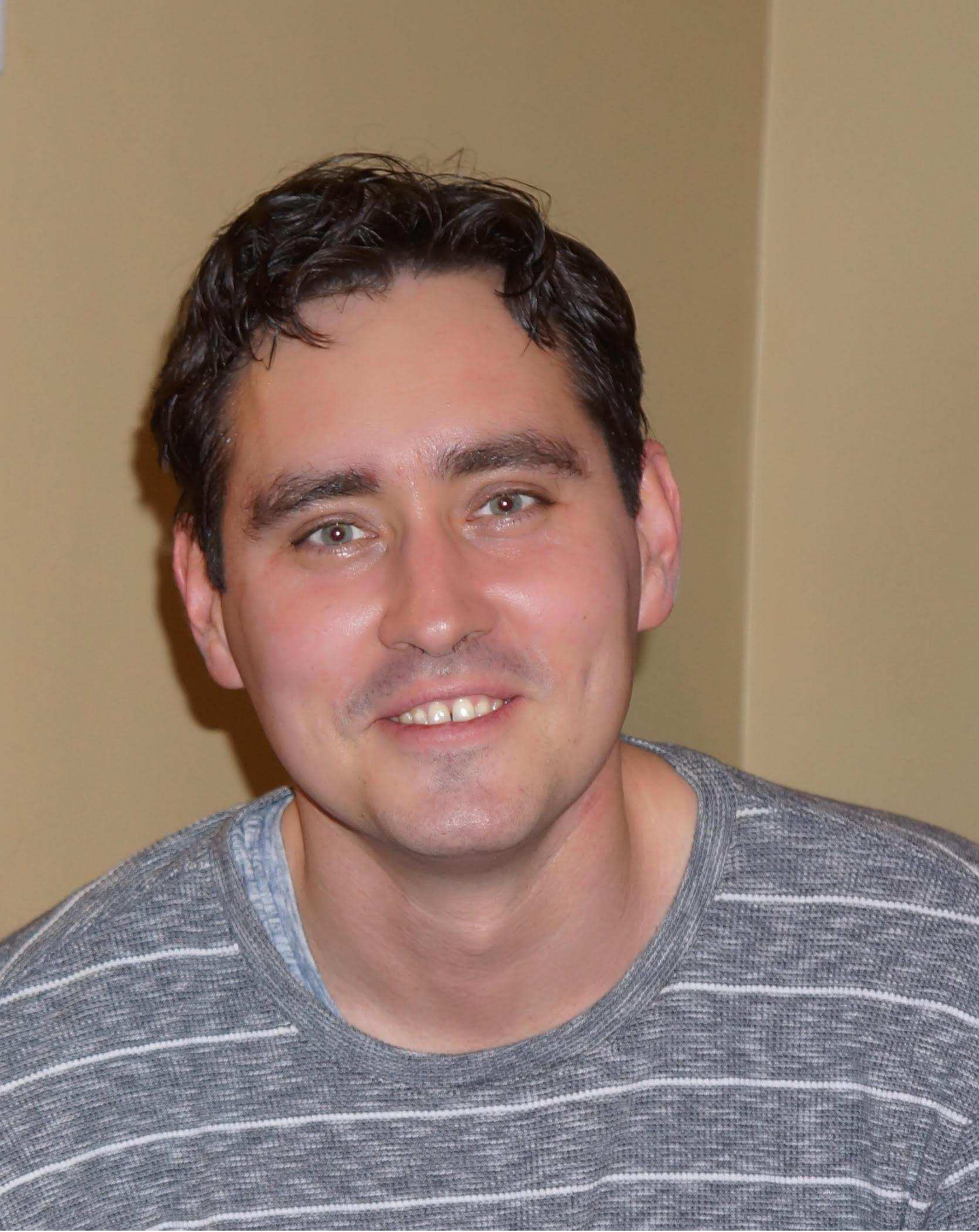}}] {{Petr Vorobev}} (M`15) received his PhD degree in theoretical physics from Landau Institute for Theoretical Physics, Moscow, in 2010.  From 2015 until 2018 he was a Postdoctoral Associate at the Mechanical Engineering Department of Massachusetts Institute of Technology (MIT), Cambridge. Since 2019 he is an assistant professor at Skolkovo Institute of Science and Technology, Moscow, Russia. His research interests include a broad range of topics related to power system dynamics and control. This covers low frequency oscillations in power systems, dynamics of power system components, multi-timescale approaches to power system modelling, development of plug-and-play control architectures for microgrids.
\end{IEEEbiography} 

\begin{IEEEbiography}[{\includegraphics[width=1in,height=1.294in]{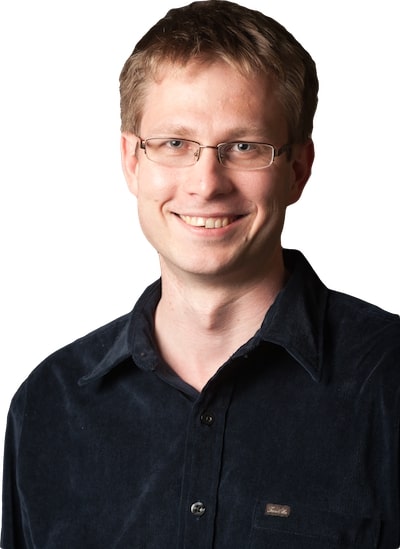}}]{{Konstantin Turitsyn}} (M`09) received the M.Sc. degree in physics from Moscow Institute of Physics and Technology and the Ph.D. degree in physics from Landau Institute for Theoretical Physics, Moscow, in 2007.  Previously, he was an Associate Professor at the Mechanical Engineering Department of Massachusetts Institute of Technology (MIT), Cambridge. Before joining MIT, he held the position of Oppenheimer fellow at Los Alamos National Laboratory, and Kadanoff-Rice Postdoctoral Scholar at University of Chicago. His research interests encompass a broad range of problems involving nonlinear and stochastic dynamics of complex systems. Specific interests in energy related fields include stability and security assessment, integration of distributed and renewable generation.
\end{IEEEbiography}
\end{document}